\documentclass[]{aa}
\usepackage[varg]{txfonts}
\usepackage{natbib}

\begin{document}

\title{A combined photometric and kinematic recipe for evaluating the nature of bulges using the CALIFA sample}

\author{J. Neumann\inst{\ref{inst1}, \ref{inst2}}
\and L. Wisotzki\inst{\ref{inst1}}
\and O.S. Choudhury\inst{\ref{inst1}}
\and D.A. Gadotti\inst{\ref{inst2}}
\and C.J. Walcher\inst{\ref{inst1}}
\and J. Bland-Hawthorn\inst{\ref{inst8}}
\and R. Garc\'ia-Benito \inst{\ref{inst3}}
\and R.M. González Delgado\inst{\ref{inst3}}
\and B. Husemann\inst{\ref{inst4}}
\and R.A. Marino\inst{\ref{inst7}}
\and I. Márquez\inst{\ref{inst3}}
\and S.F. S\'anchez\inst{\ref{inst5}}
\and B. Ziegler\inst{\ref{inst6}}
\and CALIFA collaboration}

\institute{Leibniz-Institut f\"ur Astrophysik Potsdam (AIP), An der Sternwarte 16, D-14480 Potsdam, Germany\label{inst1}
\and European Southern Observatory, Alonso de Córdova 3107, Casilla 19001, Santiago, Chile\label{inst2}
\and Sydney Institute for Astronomy, School of Physics A28, University of Sydney, NSW 2006, Australia\label{inst8}
\and Instituto de Astrofísica de Andalucía (IAA/CSIC), Glorieta de la Astronomía s/n Aptdo. 3004, E-18080 Granada, Spain\label{inst3}
\and Max-Planck-Institut für Astronomie, Königstuhl 17, D-69117 Heidelberg, Germany\label{inst4}
\and Department of Physics, Institute for Astronomy, ETH Z\"urich, CH-8093 Z\"urich, Switzerland\label{inst7}
\and Instituto de Astronom\'ia, Universidad Nacional Auton\'oma de M\'exico, A.P. 70-264, 04510 M\'exico, D.F., Mexico\label{inst5}
\and University of Vienna, Department of Astrophysics, Türkenschanzstr 17, 1180 Vienna, Austria\label{inst6}}


\abstract{

Understanding the nature of bulges in disc galaxies can provide important insights into the formation and evolution of galaxies. For instance, the presence of a classical bulge suggests a relatively violent history, in contrast, the presence of simply an inner disc (also referred to as a ``pseudobulge'') indicates the occurrence of secular evolution processes in the main disc. However, we still lack criteria to effectively categorise bulges, limiting our ability to study their impact on the evolution of the host galaxies. Here we present a recipe to separate inner discs from classical bulges by combining four different parameters from photometric and kinematic analyses: The bulge Sérsic index $n_\mathrm{b}$, the concentration index $C_{20,50}$, the Kormendy (1977) relation and the inner slope of the radial velocity dispersion profile $\nabla\sigma$. With that recipe we provide a detailed bulge classification for a sample of 45 galaxies from the integral-field spectroscopic survey CALIFA. To aid in categorising bulges within these galaxies, we perform 2D image decomposition to determine bulge Sérsic index, bulge-to-total light ratio, surface brightness and effective radius of the bulge and use growth curve analysis to derive a new concentration index, $C_{20,50}$. We further extract the stellar kinematics from CALIFA data cubes and analyse the radial velocity dispersion profile. The results of the different approaches are in good agreement and allow a safe classification for approximately $95\%$ of the galaxies. In particular, we show that our new ``inner'' concentration index performs considerably better than the traditionally used $C_{50,90}$ when yielding the nature of bulges. We also found that a combined use of this index and the Kormendy (1977) relation gives a very robust indication of the physical nature of the bulge.
}


\titlerunning{Evaluating the nature of bulges using the CALIFA sample}
\maketitle

\section{Introduction}
\label{Sect:Intro}

The traditional picture of disc galaxies consists of two main stellar components, a disc and a central spheroid - the bulge. It is a generally accepted fact that bulges play an essential role for our understanding of galaxy formation and evolution. \citet{Gadotti09} estimated that in the Local Universe bulges contribute about $28\%$ of the total stellar mass in massive galaxies. From the analysis of the stellar mass budget with the Galaxy and Mass Assembly (GAMA) survey, \citet{Moffett16} found that $15\%$ of the local total stellar mass density is distributed in S0-Sa bulges. Bulges are closely connected to the strength and length of the bar \citep[e.g.][]{Sellwood81, Aguerri09, Laurikainen09} and they are correlated with the mass of the supermassive black hole \citep[e.g.][]{Kormendy93a, Gebhardt00, Ferrarese&Merritt00, Kormendy&Gebhardt01}.

For a long time bulges were considered to be elliptical-like components embedded in an outer disc, but a significant amount of evidence has shown a dichotomy of bulges \citep[see e.g.][for a review]{Kormendy93, KormendyKennicutt04, Athanassoula05, FisherDrory15}. The bulges that fit into the traditional category of hot central elliptical-like components were henceforth called ``classical bulges'' whereas every other bulge-like, but not classical component was called ``\emph{pseudo}-bulge'' or also disc-like bulge or discy pseudobulge. Photometrically, they satisfy the definition of a bulge since they produce an excess of light over an inward extrapolation of the major disc. But they are considered to be much more like discs, e.g. they are flattened by rotation, have close to exponential light profiles and are often dominated by young stars. \citet{FisherDrory11} found that the majority of bulges in the Local Universe are in fact pseudobulges. Today we know that not only the overall bulge category, but also the pseudobulges themselves form an inhomogeneous class of objects. Morphologically, nuclear spirals, nuclear rings or nuclear bars can be part of a pseudobulge. Another sub- or equal-level category are boxy or peanut-shaped bulges. They have been shown to be the thick central parts of bars seen edge-on \citep{Kuijken&Merrifield95, Bureau&Freeman99, Bureau&Athanassoula99, Bureau&Athanassoula05, Athanassoula&Bureau99, Chung&Bureau04}. The different kinds of bulges can as well coexist \citep{Fisher08, Erwin10}. \citet{MendezAbreu14} found 7 out of 10 barred galaxies to host composite-bulges. \citet{Erwin15} predict composite-bulges to be present in at least $10\%$ of barred S0 and early-type spiral galaxies. In this paper, the term pseudobulge will be used to refer to any bulge thought to be made of disc material.

Many photometric criteria have been proposed to identify pseudobulges and classical bulges, e.g. the morphology, the concentration index, the Sérsic index \citep{Fisher08}, the Kormendy relation \citep{Kormendy77, Gadotti09} or the bulge-to-total light ratio combined with the ratio of the sizes of bulge and disc \citep{Allen06}, but none of these criteria alone can unambiguously separate the two bulge types. As a consequence, authors have used multiple criteria to improve the accuracy of bulge classification \citep[e.g.][]{KormendyKennicutt04, Fisher10, Kormendy13}.

The kinematics of bulges provided some of the earliest evidence for the dichotomy of bulges. Pseudobulges were found to be more rotationally supported as seen in the $V_m/\langle\sigma\rangle$ - $\epsilon$ diagram \citep{Kormendy82} and the central velocity dispersion was used to identify pseudobulges as low-$\sigma$ outliers from the \citet{FJ76} relation \citep{KormendyKennicutt04}. More recently there have been a few studies of kinematic bulge diagnostics that reported correlations between bulge type and radial structure of kinematics \citep{FalconBarroso06, MendezAbreu08b, MendezAbreu14, Fabricius12}. Yet, a clear quantification of the relations that they found in kinematic behaviour remains an open task. With the advent of big integral field spectroscopy (IFS) surveys, data of a new category become available to do statistically meaningful spatially resolved spectroscopy with a big sample of galaxies \citep[e.g.][]{Krajnovic11, FalconBarroso17}.

In this paper we present a combination of photometric and spectroscopic bulge indicators derived from two-dimensional analyses of the structure and kinematics of our CALIFA subsample. We use detailed growth curve measurements of the surface brightness distribution, two-dimensional photometric decompositions and kinematic maps to understand their correlation and shed light onto the nature of the bulge dichotomy. Our main aims are: 1) find a robust concentration index for bulge diagnostic, 2) use IFS data of a medium-sized sample of galaxies to investigate the bulge kinematics and 3) provide a prescription based on manifold parameters for bulge classification.

The paper is organised as follows. Section \ref{Sect:DataAndSample} describes the sample selection and data used in this work. In Sect. \ref{Sect:Methods} we describe our multiple approaches followed by Sect. \ref{Sect:Results} where we present the results. In Sect. \ref{sect:discuss} we provide a recipe for a detailed classification of bulges that we then apply to our sample, and a discussion of various aspects of our analyses. Finally, we summarise our work and main conclusions in Sect. \ref{Sect:Conclusions}. Throughout the article we assume a flat cosmology with $\Omega_\mathrm{m} = 0.286$, $\Omega_\Lambda = 0.714$ and a Hubble constant $H_0 = 69.6\, \mathrm{km\ s^{-1}\ Mpc^{-1}}$ \citep{Bennett14}.

\section{Data sources and sample selection}
\label{Sect:DataAndSample}

The Calar Alto Legacy Integral Field Area survey \citep[CALIFA, ][]{Sanchez12, Walcher14} is a large public legacy project that obtained spatially resolved spectra of about 600 local galaxies using integral field spectroscopy (IFS). The sample we use in this work is drawn from the sample of 277 galaxies that was observed in the V1200 configuration between the official start of observation in June 2010 and October 2013.

CALIFA uses the Potsdam Multi-Aperture Spectrophotometer \citep[PMAS, ][]{Roth05} instrument with the PMAS fibre package \citep[PPak, ][]{Kelz06} integral field unit (IFU) installed at the Cassegrain focus of the Calar Alto Observatory (CAHA) 3.5 m telescope in Andalucía, southern Spain. The IFU consists of a total of 382 fibres, 331 of them are object-fibres packed in a hexagonal form. Each fibre has a diameter of $2.68''$ on the sky, collecting flux from 5.7 $\mathrm{arcsec^2}$. The whole hexagonal arrangement of the object fibres covers a $74 \times 65$ $\mathrm{arcsec^2}$ field of view (FoV). The so-called CALIFA ``mother sample'' -- a pool of 939 galaxies from which the objects to observe were drawn only depending on observability -- is primarily diameter limited to ensure a good fit to the FoV of the instrument. It covers a wide range of the luminosity function and all morphological categories. We refer to \citet{Walcher14} for more details on the sample selection and characteristics.

As result of the diameter-limited aspect of the CALIFA sample $97\%$ of the galaxies are covered to more than 2 times the Petrosian half-light radius, which allows for a detailed study of the bulges and outer disc components.

The data have been reduced with the version 1.5 of the CALIFA pipeline, see \citet{Sanchez12}, \citet{Husemann13} and \citet{GarciaBenito15} for details. The final data products are two data cubes, one for the low-resolution V500 spectral setup covering the wavelength range 3745-7500 $\AA$ with a spectral resolution of 6.0 $\AA$ (full width at half maximum, FWHM) and the other for the medium-resolution V1200 setup covering the wavelength range 3650-4840 $\AA$ with a spectral resolution of 2.3 $\AA$ (FWHM). \citet{FalconBarroso17} showed that the spectral resolution of the V500 grating is not enough to accurately measure velocity dispersions below $~ 100\, \mathrm{km\,s^{-1}}$, whereas the V1200 grating allows measuring the velocity dispersion down to $~ 40\, \mathrm{km\,s^{-1}}$. We decided to use the V1200 data cubes since we are especially interested in the stellar kinematics of the bulges, which can have very low velocity dispersions if they are of discy nature.

\begin{figure}
	\resizebox{\hsize}{!}{\includegraphics{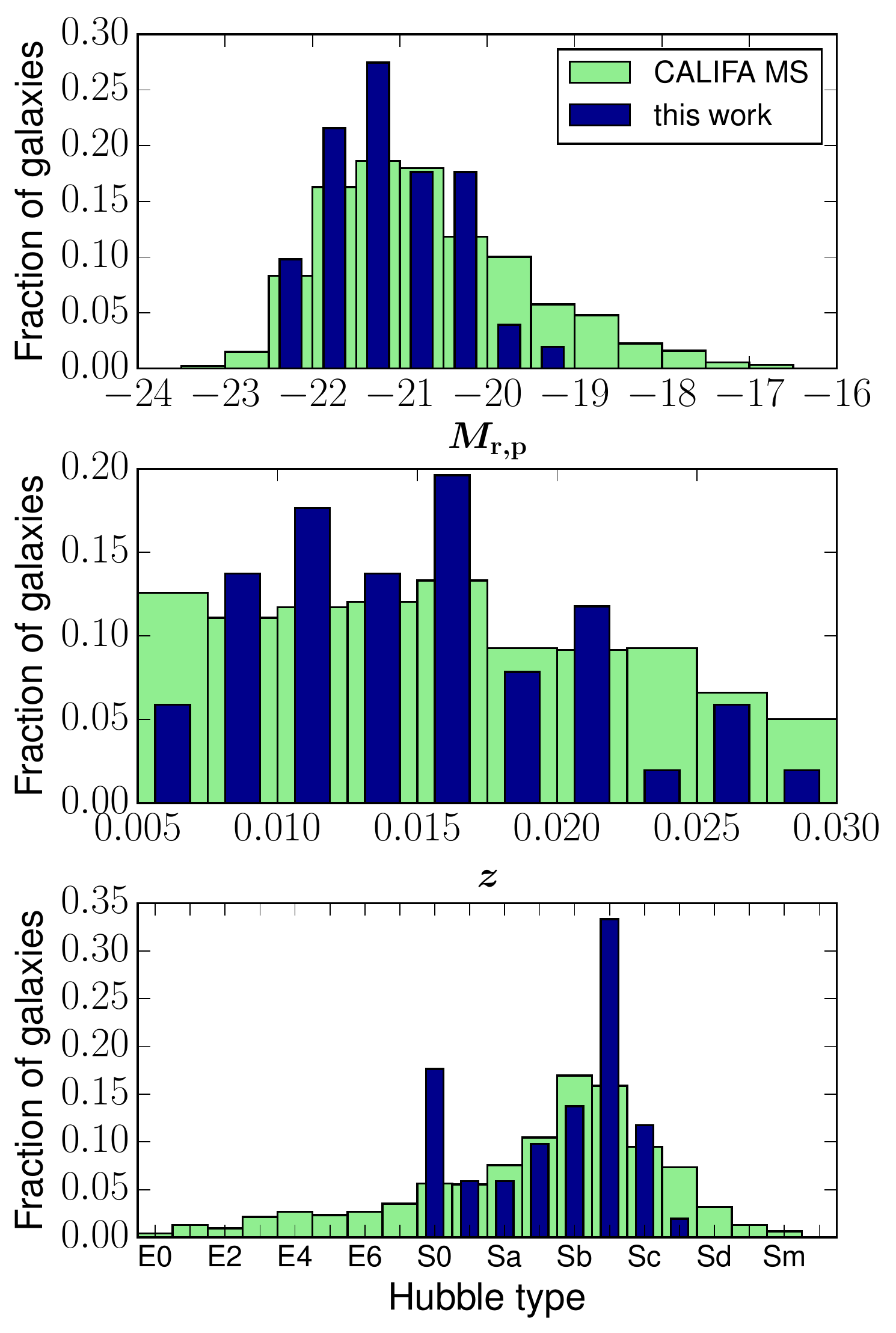}}
	\caption{Normalised histograms of absolute SDSS Petrosian $r$-band magnitudes $M_{r,\mathrm{p}}$, redshifts and morphological types of the CALIFA mother sample (MS) and the subsample we used in our analysis.}
	\label{fig:prop}
\end{figure}

The sample selection for this work was driven by the aim of investigating indicators of the bulge nature in a clean sample of undisturbed, isolated disc galaxies. From the 277 observed galaxies we selected all unbarred disc galaxies with axis ratio $b/a \geq 0.4$. Additionally, we rejected objects with problematic observational data as for example high dust obscuration, bright foreground stars or very low signal to noise (S/N). The final sample contains 45 objects, 3 of which are bulgeless galaxies for comparative purposes. Choosing unbarred galaxies as objects for our analyses is a simplification. Bars significantly influence bulge parameters like Sérsic index and bulge-to-total light ratio (B/T), if they are not properly accounted for in 2D photometric decompositions \citep{Aguerri05, Gadotti08, Salo15}. They also show kinematic features in both velocity and velocity dispersion profiles \citep[see e.g. ][]{Seidel15}. Here we tried to focus on bulge signatures by avoiding any disturbances which may originate from a bar component. In Fig. \ref{fig:prop} we show the normalised histograms of absolute SDSS Petrosian $r$-band magnitudes $M_{r,\mathrm{p}}$\footnote{For consistency with \citet{Walcher14}, we show here Petrosian magnitudes instead of total absolute magnitudes that are used in Sect. \ref{Sect:Results}.}, redshifts and morphological types of the CALIFA mother sample and the subsample we used in our analysis. We see no fundamental differences in the distribution of our subsample with the CALIFA mother sample, except that we are missing galaxies fainter than $M_{r,\mathrm{p}}=-19$ and very late-type morphological types. The morphlogical types are by our selection restricted to disc galaxies. We have additionally a significant higher fraction of S0 and Sbc galaxies, but we do not think that these differences affect the way that our analysis can be represented. We conclude that our subsample is representative of massive disc galaxies in the CALIFA mother sample. 

We first use the complete CALIFA sample for analysing some global structural parameters in Sect. \ref{sect:concentration}. Subsequently, we determine bulge parameters for our subsample of 45 disc galaxies. We additionally selected a subsample of all isolated elliptical galaxies with good observational data. We used this subsample of 26 galaxies to compare the behaviour of the bulges with respect to the scaling relations built from this sample of ellipticals.

The CALIFA sample was initially drawn from the photometric catalogue of the data release 7 \citep[DR7, ][]{Abazajian09} of the Sloan Digital Sky Survey \citep[SDSS, ][]{York00}. This ensures availability of photometric data for every CALIFA galaxy. All photometric analyses in this paper are based on the Sloan $r$-band images.

The availability of photometry from SDSS and spectroscopy from CALIFA for a large number of galaxies of all morphological types together with the large spatial coverage of the galaxies in these data makes this sample ideal for a statistical study on properties of bulges with respect to host galaxies.

\section{Methods}
\label{Sect:Methods}

\subsection{Growth curve analysis and concentration indices}

We used the results of the growth curve photometry described in detail in \citet{Walcher14} for the determination of light concentration indices. Here we just summarise the basic concept of the method. At first step masks were produced by a combination of an automatic algorithm with SExtractor \citep{Bertin&Arnouts96} and individual inspection by eye. The position angle (PA) and axis ratio ($b/a$) values were derived from second order moments of the SDSS $r$-band light distribution. The growth curve was then measured within concentric elliptic rings with increasing major axes and fixed $b/a$ and PA.The most delicate and important part is the accurate determination of the sky background and the edge of the galaxy in the presence of sky gradients and incomplete masks. The edge of the galaxy was determined as the major axis at the middle of the current ring where the flux profile slope becomes non-negative, the sky value as the mean of the values within that ring. This method was shown to be sufficiently robust.

From these growth curves it is possible not only to determine the half-light semi-major axes of the galaxies but any kind of ellipse that encloses a certain percentage of the total light of the galaxy. We denote from here on the semi-major axis as approximation of the radius $r_k$ that encircles k percent of the total light, e.g. $r_{20}$ is the radius that encloses 20 percent of the light. The ratio of one of these radii divided by another in any combination can then be used as a concentration index of the galaxy. After exploring a wide range of different options we have chosen $C_{20,50} = r_{20}/r_{50}$ and $C_{50,90} = r_{50}/r_{90}$ as an inner and outer concentration index, respectively. Details on the motivation for that choice and results are presented in Sect. \ref{sect:concentration}.

\subsection{Two-dimensional image fitting}

Two-dimensional photometric decomposition has become a widely used technique for deriving the structural parameters of galaxies. Multiple codes have been developed to perform image decomposition, such as \emph{GIM2D} \citep{Simard02}, \emph{GALFIT} \citep{Peng02, Peng10}, \emph{BUDDA} \citep{Souza04} and \emph{GASP2D} \citep{MendezAbreu08a, MendezAbreu14}. 2D image fitting can be very fast in determining in an automatic way single component parameters such as disc scale length and central surface brightness, but it becomes highly complex and sensitive to initial parameter guesses when fitting in parallel multiple functions to the data. We used \emph{IMFIT} by \citet{Erwin15a}\footnote{http://www.mpe.mpg.de/$\sim$erwin/code/imfit/} to fit single Sérsic functions to the whole set of 939 galaxies from the CALIFA mother sample. These fits gave us a global Sérsic index for each galaxy that we denote as $n_\mathrm{g}$.

In addition to that, we chose to perform a two-component bulge-disc decomposition for our sample of 45 galaxies to derive Sérsic indices $n_\mathrm{b}$ of the bulge component only. We used a Sérsic function for the bulge and an exponential for the disc. In cases of a Type II \citep{Freeman70} or Type III disc \citep{Erwin05} we either used the \emph{BrokenExponential} function of \emph{IMFIT} or we restricted our fit to the central disc component by masking out the outer region of the galaxy. This is a valid approach since we are interested in the central component only. We also checked our sample for evidence of a nuclear component and found no need to fit a central point source. Except for one galaxy that hosts a low-ionization nuclear emission-line region (LINER), none of the objects in our sample hosts active galactic nulei (AGNs). From the best fit model parameters we derived bulge-to-total light ratios ($B/T$).

Many image fitting codes provide formal uncertainties on the parameter estimates from the Levenberg-Marquardt minimisation technique. In \emph{IMFIT} there is additionally a \emph{bootstrap} re-sampling option that can be used. The relative uncertainties that we estimated using the bootstrap option for our analysis are on the order of a few per cent. \citet{Gadotti09} found uncertainties of bulge, disc and bar parameters to be in the range of 5-20\% using a different statistical approach with \emph{BUDDA}.  However, it has repeatedly been shown that all estimates of statistical errors should be considered underestimates of the true uncertainty of the parameters \citep[e.g.][]{Haeussler07, MendezAbreu08a, Gadotti09, Erwin15a}. One relevant source of uncertainty is the human factor when it comes to select the best model to fit to the data. This is very difficult to account for in a proper error estimation. Additionally, multi-component fits are sensitive to input parameters the more complex the galaxy structure becomes. We therefore think that these error estimates are not representative and we chose not to show errorbars for the structural parameters on the individual plots in our paper.

\begin{figure*}
\sidecaption
	\includegraphics[width=12cm]{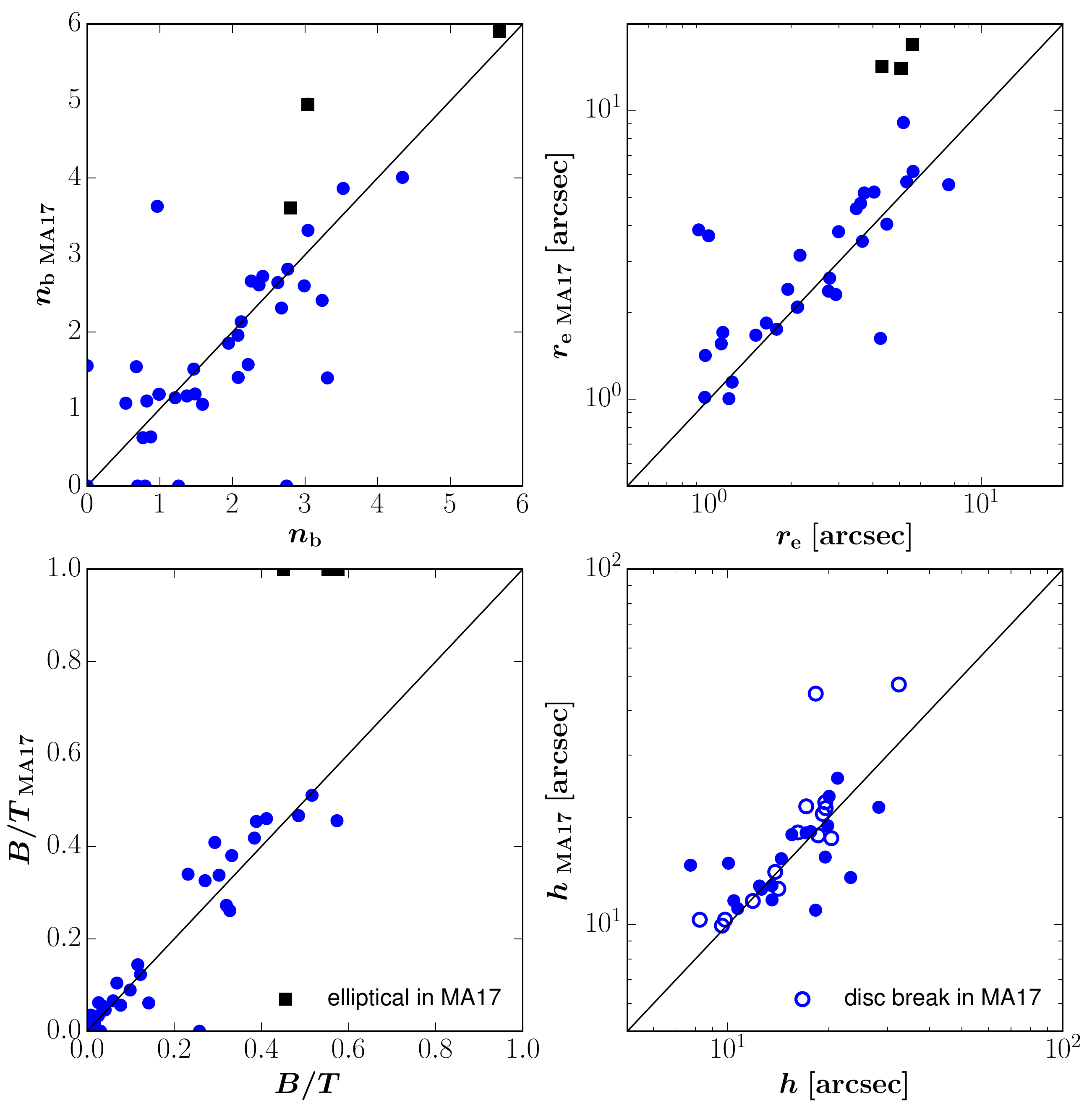}
	\caption{Comparison between our decomposition parameters and those obtained from the \emph{GASP2D} $r$-band decompositions from \citet[annotated as MA17]{MendezAbreu17}. The two samples have 40 galaxies in common. The left panels show the Sérsic index of the bulge (top) and the bulge-to-total luminosity (bottom). The right panels show the bulge effective radius (top) and the disc scale length (bottom). Galaxies with disc breaks in the decompositions of MA17 are marked by empty circles. In these cases we plot the scale length of the inner disc. Objects with $n_\mathrm{b}=0$ in our decomposition are galaxies that were classified as bulgeless. Objects with $n_\mathrm{b\ MA17}=0$ in their decomposition were either classified as bulgeless or as having an unresolved bulge. In  the latter case, they were modeled by them with a nuclear point source instead. The three objects that have $B/T =1$ were classified by them as purely spheroidal. There is an overall agreement between both decompositions with moderate scatter.}
	\label{fig:cross}
\end{figure*}

Recently, \citet{MendezAbreu17} (from here on denoted as MA17) published results from photometric decompositions of 404 CALIFA galaxies in the $g$, $r$ and $i$ SDSS images using the \emph{GASP2D} code. In Fig. \ref{fig:cross} we compare our best-fit parameters with the results they obtained for the $r$-band images. 

The two samples have 40 galaxies in common. We find a relative good agreement between both decompositions. Three of the galaxies were classified in MA17 as purely spheroidal and have therefore $B/T_\mathrm{MA17}=1$, and comparitively high $n_\mathrm{b\ MA17}$ and $r_\mathrm{e\ MA17}$. The other outliers from the one-to-one relation correspond to either dusty or moderately inclined galaxies or to objects with a more complex structure. Given the overall similarity of the results of both analyses, we will use our decomposition parameters throughout the paper. 

\subsection{Definition of bulge radius}
\label{sect:bulgeradius}

We define the ``bulge radius'' as the radius of the ellipse that encloses $90\%$ of the light of the bulge component. This radius is determined by numerical integration of the best fit model Sérsic function from the two-component decomposition. We denote this radius throughout the paper as $rb_{90}$. This radius is not a demarcation between bulge-dominated and disc-dominated region, but a limit of the bulge extent. The choice was made in order to trace the whole region of bulge influence. The decision was to use a radius as large as possible, but only as long as the bulge is still significant.

\subsection{Velocity dispersion measurement}

One key point of this work is to combine photometric bulge indicators with spectroscopic approaches. The CALIFA IFU data offer the great possibility of studying the stellar kinematics in a 2D plane over a large extent of the galaxies. We used the data from the medium resolution V1200 spectral setup to create velocity dispersion profiles from azimuthally averaged stellar kinematic maps for all 45 galaxies of our sample. The procedure is as follows:

We first binned the data cubes spatially using the 2D Voronoi binning method of \citet{Cappellari03} to achieve a constant S/N per spatial bin of 5. This allows for a sufficiently accurate measurement of the velocity and to maintain at the same time enough spatial resolution elements. In the case of 5 galaxies we had to apply a higher S/N of 10 to get reliable velocities. This S/N limit is for the velocity measurement only, not the velocity dispersion. Since we are determining only the shift of the spectrum over a wavelength range of 700 $\AA$ with good spectral resolution, we consider this S/N limit sufficient. When calculating the noise, we applied weights to the errors in order to take into account the effect of correlated noise of nearby spaxels \citep{Husemann13}.

We then estimated the velocities for each bin by fitting model template spectra from the full INDO-US template library \citep{Valdes04} to the observed spectra using the code \emph{PyParadise} \citep{Husemann16} which is an extended Python version of \emph{paradise} \citep{Walcher15}. We refer the reader to these references for the details of the algorithm. We used stellar absorption fitting only, since we are only interested in the stellar kinematics. We limited the fit to the wavelength region 4100-4800$\,\AA$ and masked out strong emission lines. Prior to the fit the stellar templates are smoothed with a 2.3$\,\AA$ (FWHM) kernel to match the wavelength resolution of the observed CALIFA data. In the \emph{PyParadise} run, a Markov-Chain-Monte-Carlo (MCMC) algorithm is used to determine the velocities and related uncertainties.

The next step was to correct every spaxel of the original unbinned data cube to rest-frame and then to the systemic velocity of the galaxy. Afterwards, we binned the cube radially in elliptical rings of 1 pixel widths which corresponds to $1\,\arcsec$ on the sky. Again, we carefully considered the correlated noise during the calculation of the error values. At this point each radial bin is represented by one spectrum.

Finally, we run \emph{PyParadise} again for each radial bin to estimate the velocity dispersions and associated uncertainties.

\begin{figure}
	\resizebox{\hsize}{!}{\includegraphics{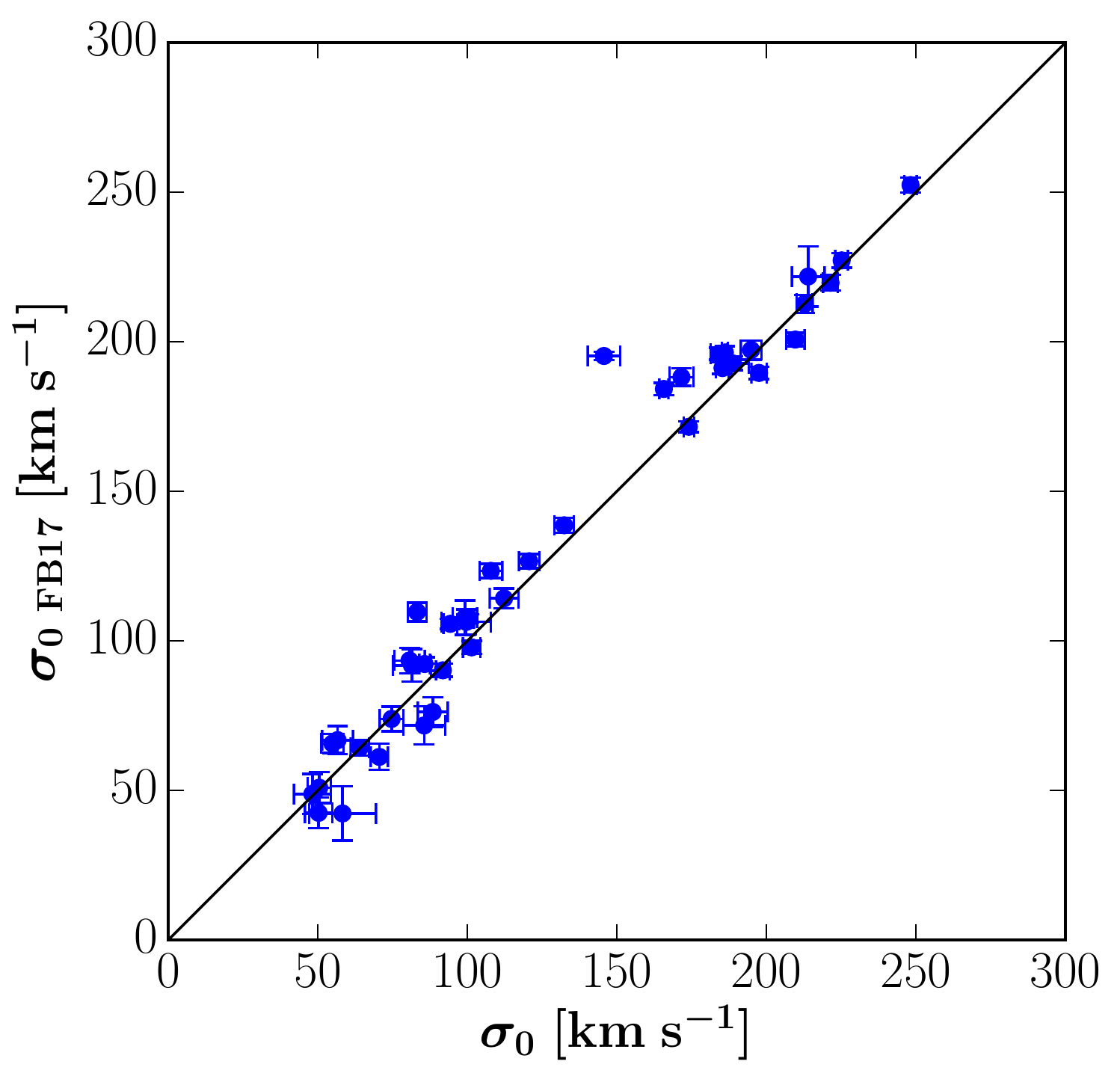}}
	\caption{Comparison between our central velocity dispersion measurements and those obtained from \citet[annotated as FB17]{FalconBarroso17}. Overall the results are in good agreement with the exception of one galaxy.}
	\label{fig:comp_kine}
\end{figure}

In Fig. \ref{fig:comp_kine} we compare our measurements of the central velocity dispersion with the results from \citet{FalconBarroso17} who performed a detailed kinematic analysis of a sample of 300 CALIFA galaxies. Both analyses have 43 galaxies in common. The results are in good agreement with only one outlier that has a significant higher central velocity dispersion in their measurement. We also conducted a case-by-case comparison between our radial velocity dispersion profiles and theirs. We did not find major differences between both results.

\section{Results}
\label{Sect:Results}

In this section we present our findings from the different approaches to characterise bulges and show correlations between the parameters we measured. We highlight the advantages in using the concentration index $C_{20,50}$ and put special effort to combine the photometric approaches with the kinematic measurements. We use both the bulge Sérsic index $n_\mathrm{b}$ and the concentration index $C_{20,50}$ to separate groups of bulges in the plots. However, we carefully point out that this is by no means meant to be a final classification into classical bulge and pseudobulge. In Sect. \ref{sect:recipe} we give a recipe using a combination of various parameters for that purpose. The results of our analyses are summarised in Table \ref{tbl:alldata} in Appendix \ref{apx:alldata}.

\subsection{Light concentration}
\label{sect:concentration}

\begin{figure}
	\resizebox{\hsize}{!}{\includegraphics{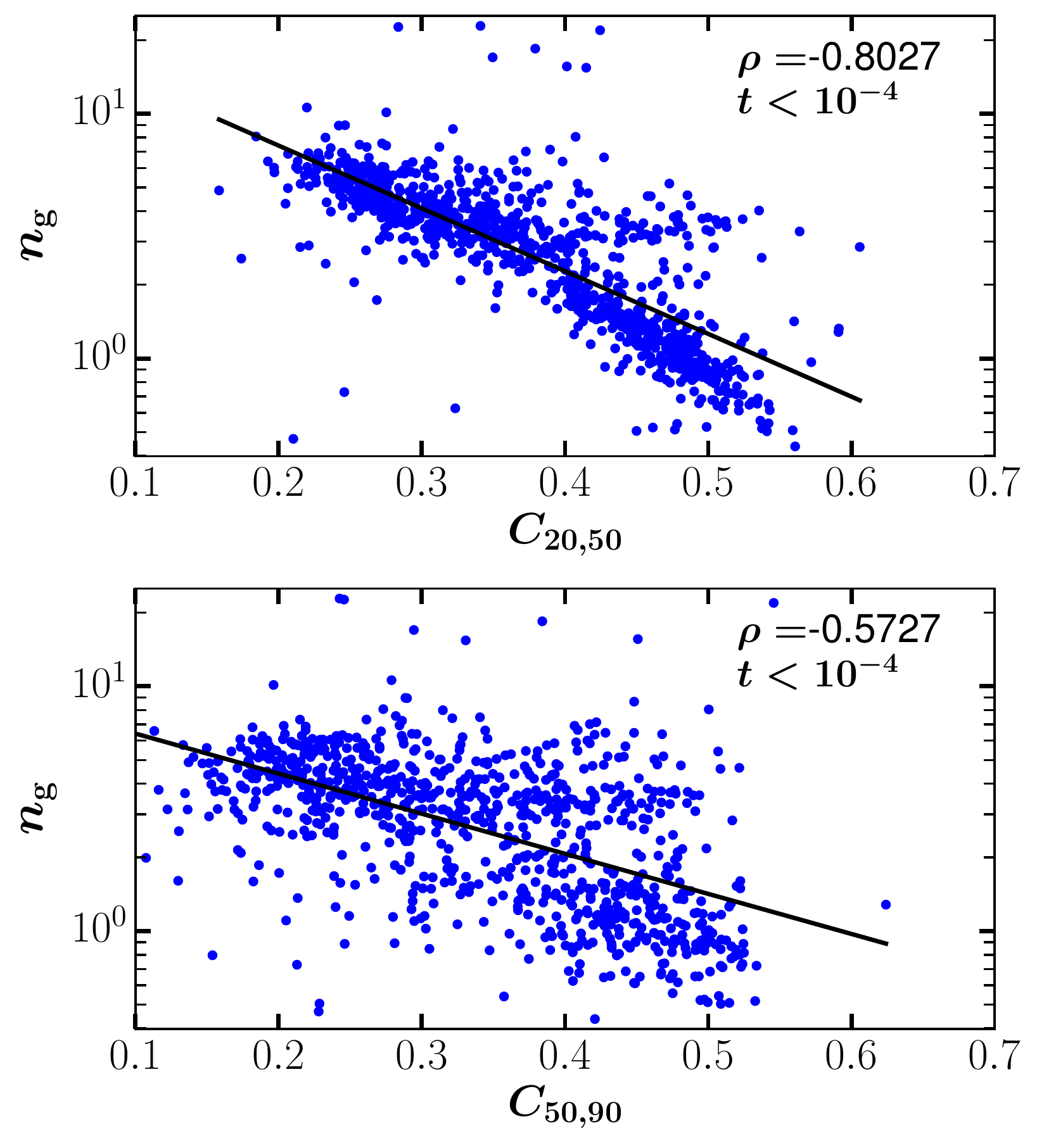}}
	\caption{Anti-correlation between log($n_\mathrm{g}$) and $C_{20,50}$ (upper panel) and $C_{50,90}$ (lower panel). The Spearman's rank correlation coefficient $\rho$ and the value of the null hypothesis significance test $t$ are given in the figure.}
	\label{fig:sidx_cidx}
\end{figure}

In Fig. \ref{fig:sidx_cidx} we present concentration indices from the growth curve measurement and global Sérsic indices from the image fitting for the complete set of 939 galaxies that compose the CALIFA mother sample. The global Sérsic index $n_{\mathrm{g}}$ is the index obtained from single Sérsic function fits to the galaxies.

We find a very tight correlation of the logarithm of $n_{\mathrm{g}}$ with $C_{20,50}$ with a Spearman's rank correlation coefficient of $\rho = -0.8$. A relation between both parameters is expected and has been shown, since they are both estimators of the steepness of the surface brightness profile \citep[e.g.][]{Trujillo01, Andrae11, Ferrari15}. Our point is to show the differences between ``outer'' and ``inner'' concentration indices. Previous studies \citep[e.g.][]{Gadotti09} have shown that the Petrosian index $R_{90}/R_{50}$ correlates with Sérsic index but with considerable scatter. In the bottom panel we see that there is indeed much more scatter in the relation between our measurements of $n_\mathrm{g}$ and $C_{50,90}$. This clearly favours the usage of the $C_{20,50}$ concentration index as a discriminator between bulge-dominated and disc-dominated galaxies.

It is worth noting that at least part of the reduced scatter using $C_{20,50}$ instead of $C_{50,90}$ might be caused by a smaller uncertainty in the determination of the concentration index. While the radii $r_{20}$ and $r_{50}$ are located on the steeper part of the growth curve, $r_{90}$ is likely to be on the shallower part, where smaller errors in the flux measurement lead to larger uncertainties in the determination of the radius.

\begin{figure}
	\resizebox{\hsize}{!}{\includegraphics{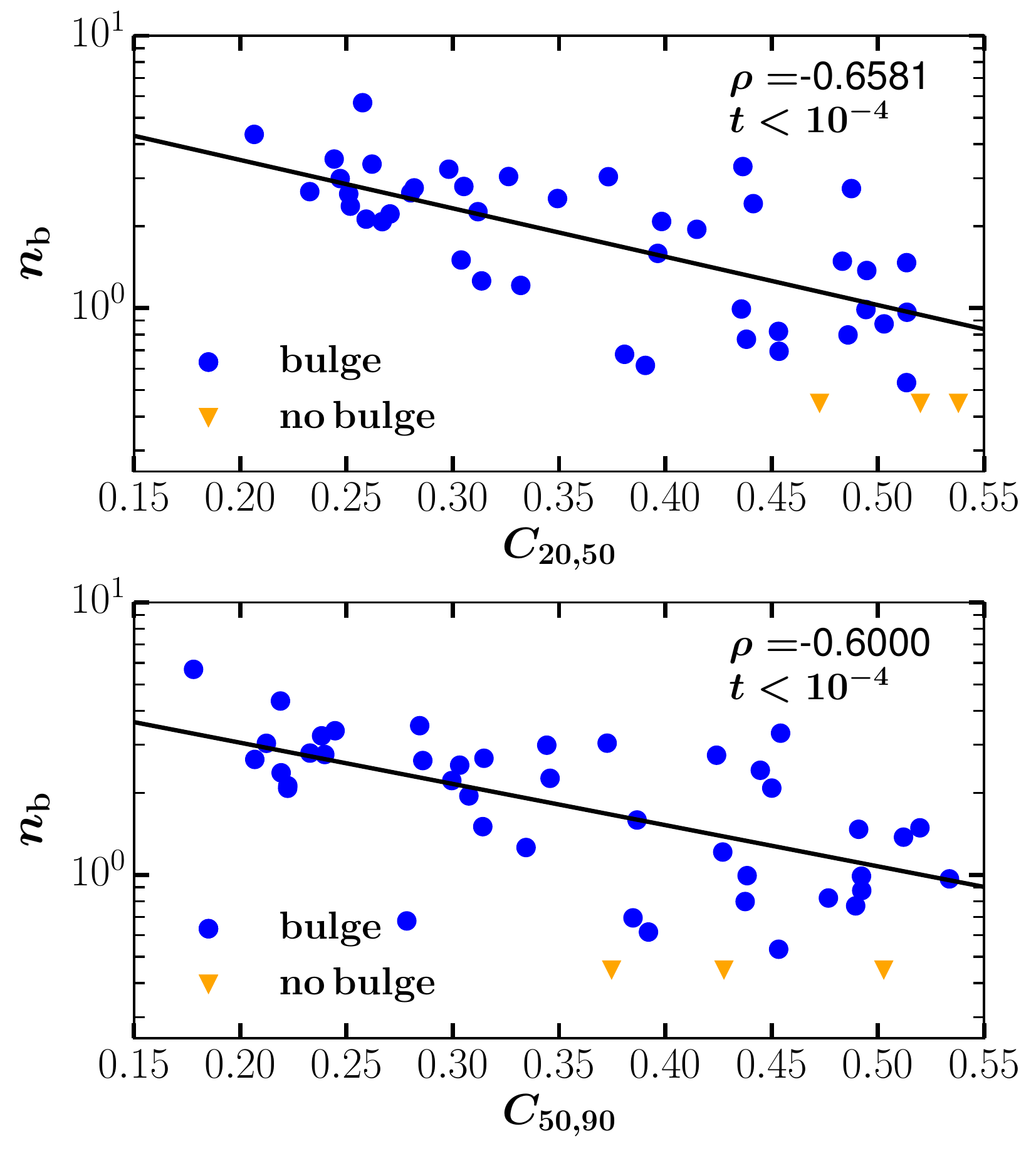}}
	\caption{Anti-correlation between log($n_\mathrm{b}$) and $C_{20,50}$ (upper panel) and $C_{50,90}$ (lower panel). The Spearman's rank correlation coefficient $\rho$ and the value of the null hypothesis significance test $t$ are given in the figure. Bulgeless galaxies are marked by orange triangles for comparison. The y-axis value of the bulgeless galaxies is set fixed to some randomly chosen value for displaying purposes.}
	\label{sbidx_cidx}
\end{figure}

We produced the same plots, but with the Sérsic index from the bulge component only for our much smaller sample of 45 galaxies. This is shown in Fig. \ref{sbidx_cidx}. The two plots in this figure suggest that $C_{20,50}$ indicates how much the light or mass of the bulge in a given galaxy is centrally concentrated and it does so better than $C_{50,90}$. The difference is less noticeable than in Fig. \ref{fig:sidx_cidx}, but still existent (The upper panel shows a Spearman's rank of $\rho = -0.66$, while for the lower panel we measure $\rho = -0.60$). It is also striking that bulgeless galaxies cover a wide range of concentration values when one uses $C_{50,90}$, but are confined to low values when one uses $C_{20,50}$, more in line with the fact that these galaxies have no bulges. This also favours our use of $C_{20,50}$ over $C_{50,90}$ as a reliable bulge parameter.

\begin{figure}
	\resizebox{\hsize}{!}{\includegraphics{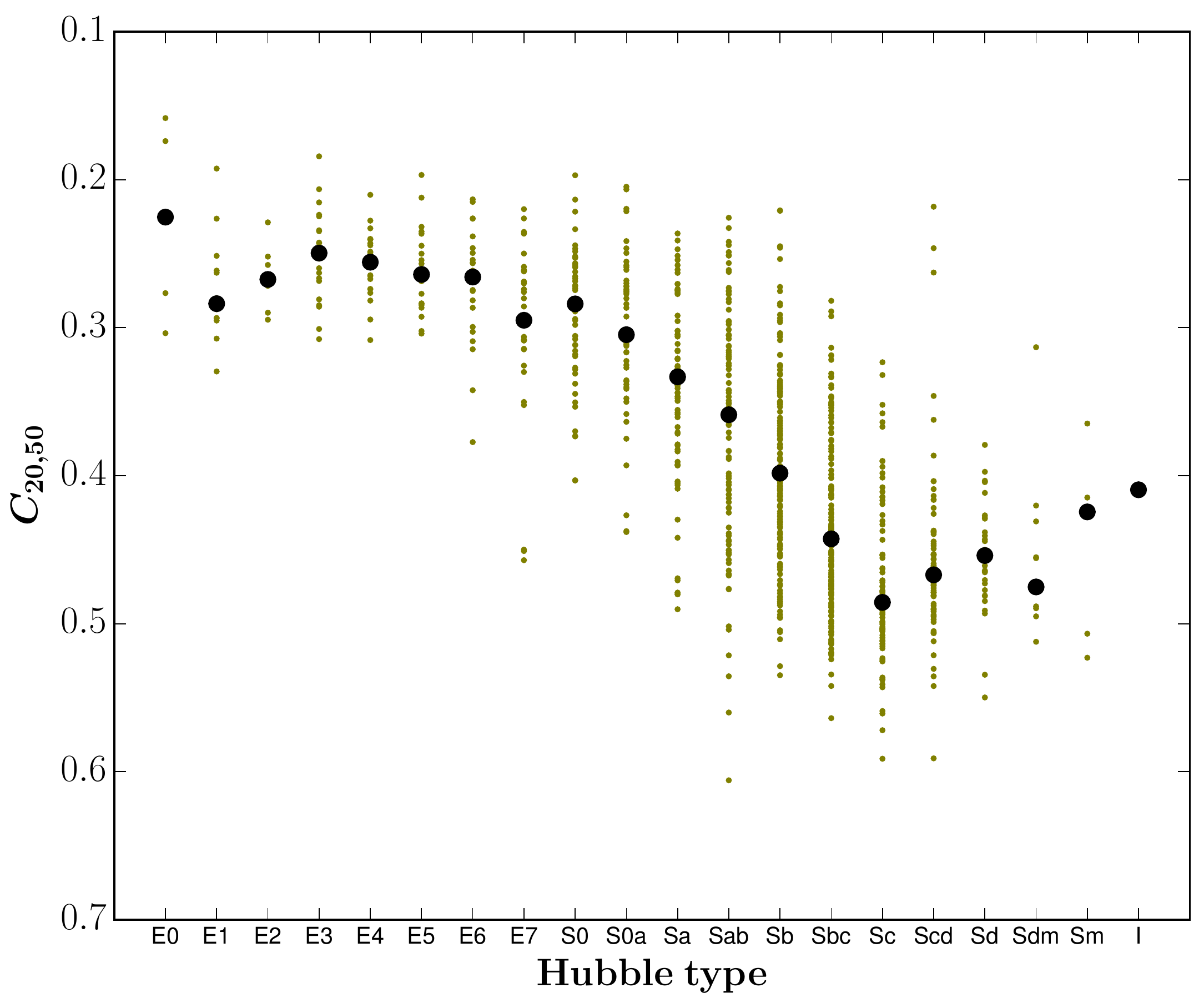}}
	\caption{Relation between $C_{20,50}$ and morphological type for the complete CALIFA mother sample. Median values for each type are marked by big black dots.}
	\label{fig:cidx_htype}
\end{figure}

Pseudobulges were suggested to be more frequent in late-type galaxies whereas classical bulges are more often found in early types \citep[e.g.][]{KormendyKennicutt04}. It is therefore interesting to investigate how our $C_{20,50}$ relates to the morphological type. Fig. \ref{fig:cidx_htype} shows this relation for the complete CALIFA mother sample. We observe a relatively flat distribution for all elliptical galaxies followed by a continuous decrease in concentration (higher values in $C_{20,50}$) from S0 to Sc, where we find the minimum in the distribution of the concentration, and finally there is a slight increase towards very late types. These results are in line with the expectation given the classification criteria of the morphological types. The increase in concentration for very late types might be surprising, however, the sample statistics are getting lower for these categories. We would like to point out the median concentration for Sb galaxies $\langle C_{20,50}\rangle = 0.398$. \citet{KormendyKennicutt04} found a sharp transition between the ocurrence of classical bulges and pseudobulges at Hubblte type Sb. Completely independently, we decided to use $C_{20,50}=0.4$ as demarcation between bulges and inner discs in Sect. \ref{sect:recipe} based only on the correlation with other classification criteria. The almost perfect agreement reinforces our decision to use $C_{20,50}=0.4$ for separating inner discs from classical bulges.

We caution the reader to be aware of the difference between concentration indices derived from growth curve measurements and those extracted from within Petrosian radii. In Appendix \ref{apx:petro} we show the relation between both approaches and provide a conversion factor between our concentration index $C_{20,50}$ and the associated Petrosian concentration.

\subsection{Structural Properties}

\begin{figure}
	\resizebox{\hsize}{!}{\includegraphics{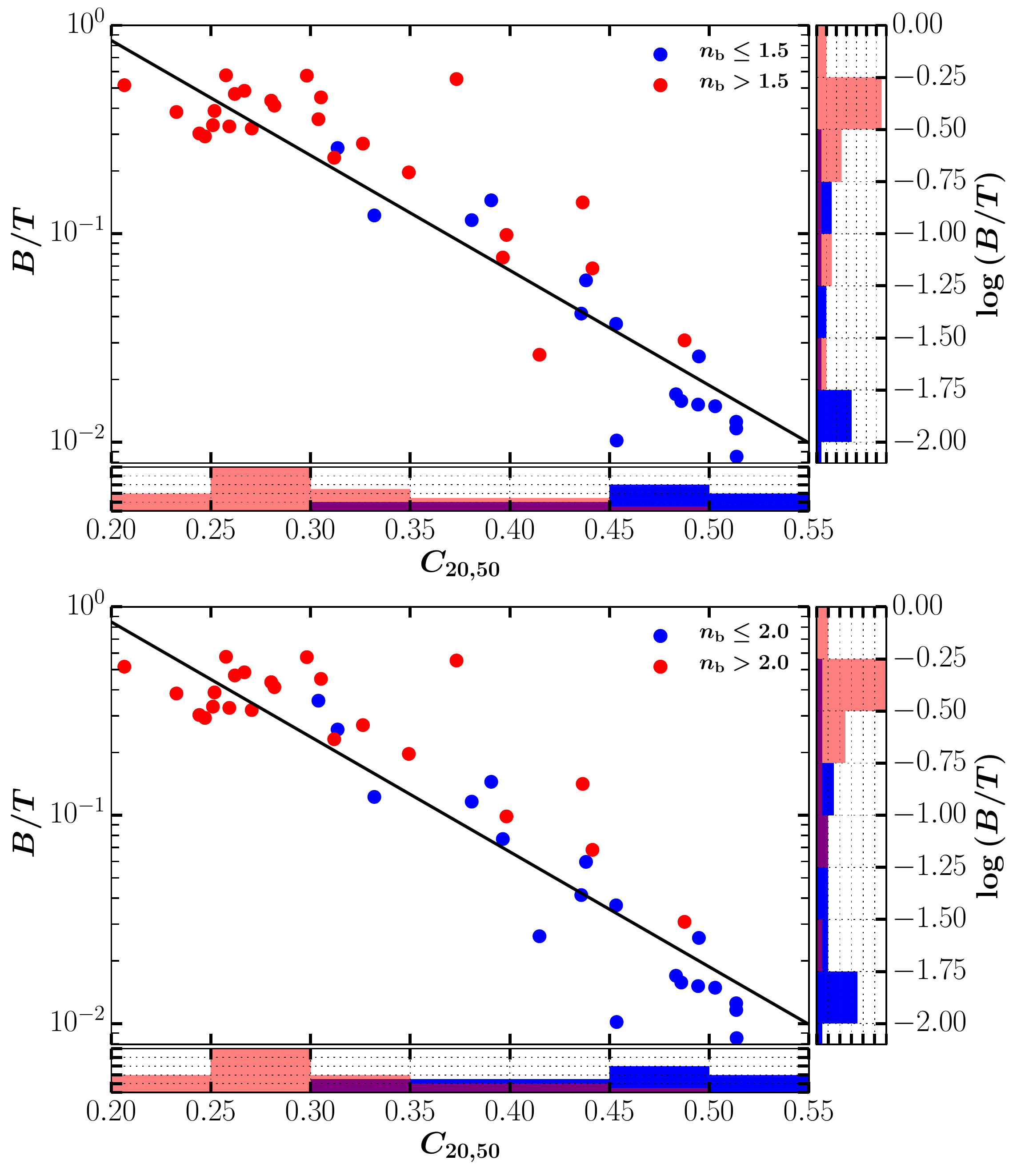}}
	\caption{Anti-correlation between log($B/T$) and $C_{20,50}$. Blue and red colours correspond to low- and high-$n_\mathrm{b}$ values of the associated bulge component. The upper and lower panels compare two different boundaries chosen to separate bulge Sérsic indices. The histograms on the right side show the distribution of the sample in equidistant bins in logarithmic space of $B/T$. The histograms on the lower side show the distribution of the concentration index.}
	\label{B_D}
\end{figure}

In this subsection we present bulge and disc component parameters derived from the 2D image decomposition. We show in Fig. \ref{B_D} the relation between bulge-to-total light ratio ($B/T$) and concentration index $C_{20,50}$. A clear correlation can be seen in the sense that more bulge-dominated galaxies have higher concentrations, reflected as lower values of $C_{20,50}$. This is what one would expect from a theoretical point of view, but the strength of the correlation with a Spearman's rank of $\rho=-0.86$ is surprising and encourages even more the use of $C_{20,50}$ as preferable concentration index.

Additionally, we divided the objects in the upper panel into low-$n_\mathrm{b}$ ($n_\mathrm{b} \leq 1.5$) and high-$n_\mathrm{b}$ ($n_\mathrm{b} > 1.5$) galaxies and we observe that the low-$n_\mathrm{b}$ galaxies do not populate the region of $C_{20,50} < 0.3$ and $\log(B/T) > -0.5$ and only one outlier has $\log(B/T) > -0.75$. This is an indication that pseudobulges correspond to low $B/T$ fractions.

The lower panel shows the same plot, but the galaxies are separated by $n_\mathrm{b} = 2$, a value that has more commonly been used in the literature for bulge separation. With this approach we see a slightly higher fraction of low-$n_\mathrm{b}$ galaxies in the region of high concentration and high bulge fraction, which is expected to be populated by classical bulges. A further comparison of both choices with all following bulge diagnostics showed that the bulges of our sample with a Sérsic index between 1.5 and 2.0 are more likely to be classical bulges. We therefore choose $n_\mathrm{b} = 1.5$ over $n_\mathrm{b} = 2$ as boundary between low- and high-$n_\mathrm{b}$ galaxies for the rest of the paper.

\subsection{Kormendy relation}

The \citet{Kormendy77} relation is a relationship between effective radius $r_\mathrm{e}$ and mean effective surface brightness $\langle\mu_\mathrm{e}\rangle$ that has been found for elliptical galaxies. It has been used by \citet{Gadotti09} and \citet{Fisher10} to study the location of bulges in a projection of the fundamental plane \citep{Djorgovski87, Dressler87}. The authors of both works found that pseudobulges tend to have lower surface brightness than classical bulges or elliptical galaxies of similar sizes. \citet{Gadotti09} even favoured this criterion for the identification of pseudobulges over the Sérsic index or the bulge-to-total light ratio.

\begin{figure}
	\resizebox{\hsize}{!}{\includegraphics{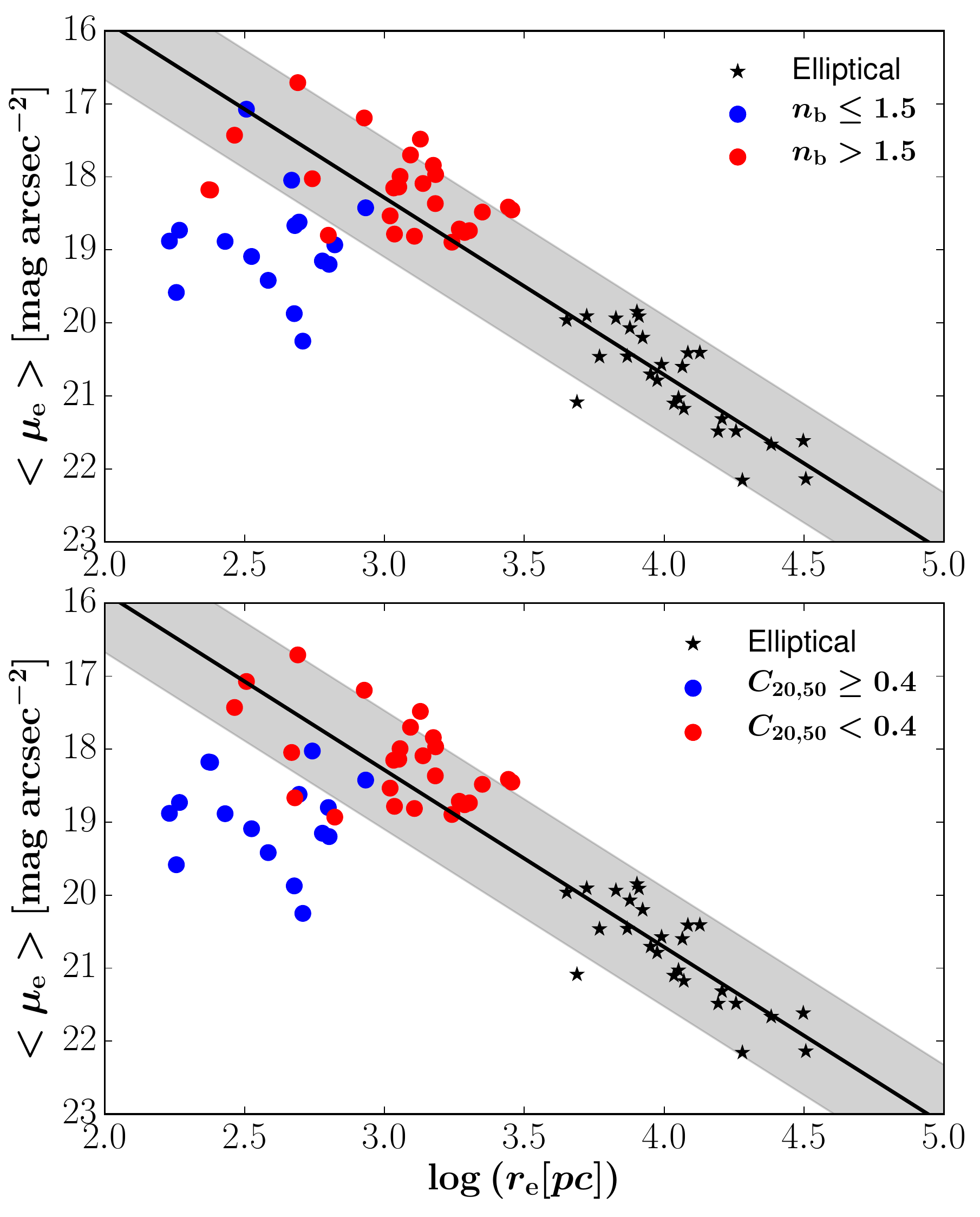}}
	\caption{Mean effective surface brightness within the effective radius vs. logarithm of the effective radius for bulges and ellipticals. The Kormendy relation is represented by a fit to the elliptical galaxies. The grey shaded region marks the $\pm 2 \sigma$ boundaries. \emph{The only difference between the two panels is the colour coding. The upper panel shows a separation of the bulges by the Sérsic index, the lower panel shows a separation according to the concentration index.} The resemblance of both plots demonstrates the equal usability of the two parameters $n_\mathrm{b}$ and $C_{20,50}$ for bulge classification.}
	\label{fig:Korm}
\end{figure}

In Fig. \ref{fig:Korm} we present the Kormendy relation for our sample. We observe a relatively clear separation of three independent groups: Ellipticals, high-$n_\mathrm{b}$ and low-$n_\mathrm{b}$  bulges in the upper panel and ellipticals, low and high concentration galaxies in the lower panel. Nearly all bulges that would be classified as classical based on the Sérsic index and concentration are located within the $2$-$\sigma$ boundaries of the relation found for elliptical galaxies and allmost all pseudobulges are below the relation. The overlap between the two types of bulges is marginal. A co-location with the Kormendy relation demonstrates the similarity of the structure of these bulges with elliptical galaxies. We see that both classifications agree very well with the Kormendy relation criterion. In addition, Fig. \ref{fig:Korm} also indicates that using $n_\mathrm{b}$ and $C_{20,50}$ for classifying bulges should yield statistically similar results, a point worth noting, given that $C_{20,50}$ is much more straightforward to derive.

The results confirm the value of the Kormendy relation for bulge diagnostics and we use it in Sect. \ref{sect:recipe} for the overall classification. 

\subsection{Faber-Jackson relation}

\begin{figure}
	\resizebox{\hsize}{!}{\includegraphics{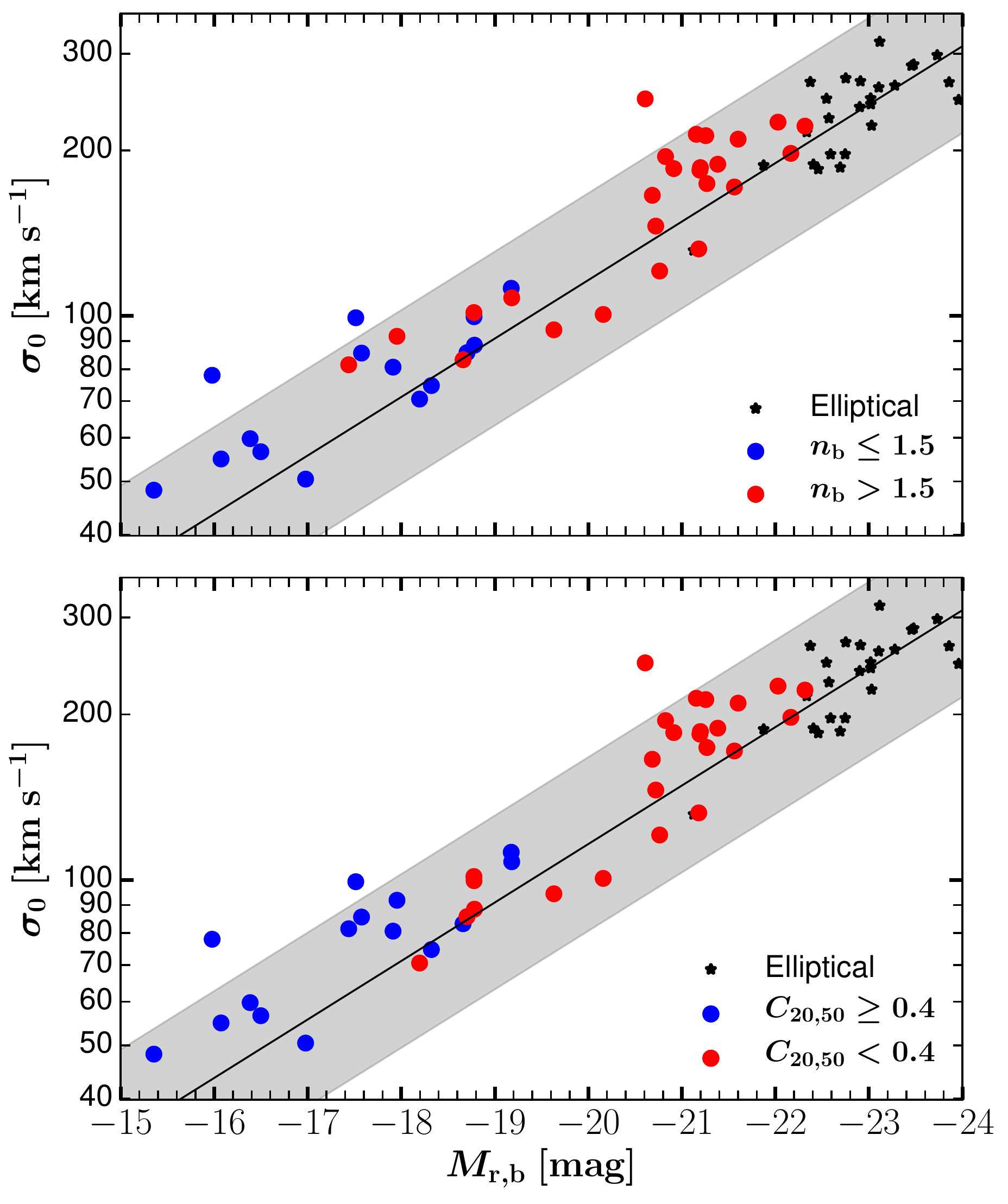}}
	\caption{Central velocity dispersion vs. absolute $r$-band magnitude for bulges and ellipticals. The Faber-Jackson relation is represented by a fit to the elliptical galaxies. A grey shade points out the $\pm 3 \sigma$ boundaries. Blue and red colours show the bulge separation by $n_\mathrm{b}$ (upper panel) and $C_{20,50}$ (lower panel).}
	\label{FJ}
\end{figure}

Some bulges of late-type galaxies have been reported to be low-$\sigma$ outliers from the \citet{FJ76} relation \citep{KormendyKennicutt04}. In Fig. \ref{FJ} we present the relation of central velocity dispersion $\sigma_0$ with absolute $r$-band magnitude $M_{r,\mathrm{b}}$ of the bulge component for our sample and we added again the subsample of elliptical galaxies from CALIFA to fit the Faber-Jackson relation for ellipticals ($L \propto \sigma_0^{\gamma}$). In the upper panel we distinguish again between low-$n_\mathrm{b}$ and high-$n_\mathrm{b}$ galaxies, whereas in the lower panel we divide the galaxies based on the concentration index. We define the central velocity dispersion as the mean velocity dispersion within $1''$. This is calculated using the binned radial profiles of $\sigma$.

We do not observe any low-$\sigma$ outliers from the Faber-Jackson relation. In fact, all bulges -- independent of their concentration or their shape of the surface brightness profile -- do align with the elliptical galaxies within the normal range of scatter. This means that either all bulges have a physical similar structure to the elliptical galaxies and are not inner discs, or the Faber-Jackson relation is not a good instrument to separate inner discs from classical bulges. We believe that the latter is the case. It has been reported that the spread in this relation is usually large and a co-location with the elliptical galaxies does not mean that the object is a classical bulge \citep[e.g.][]{FisherDrory15}. Moreover, for almost all bulges the Sérsic index and the concentration index agree very well with the concept of having a different physical nature as seen in the Kormendy relation. This indicates that the central velocity dispersion is probably more related to the total mass of the galaxy and not the central component alone. We should instead analyse the radial distribution of the velocity dispersion.

\subsection{Velocity dispersion gradient}
\label{sect:result_sigma}

When measuring the velocity dispersion profile it is important to keep in mind that it can be affected in the central region by limits in the spatial resolution. The innermost values would be smeared out to a flat profile. \citet{Gadotti08} showed that the structural properties of galaxies can be reliably determined, if the effective radius is larger than $0.8$ times the half width at half maximum (HWHM) of the point spread function (PSF). This criterion was estimated for photometric approaches by fitting two-dimensional galaxy images. In spite of that, it should be adaptable for spatially resolved kinematic analyses. The median PSF FWHM of our galaxies is $\approx 2.4''$ \citep{GarciaBenito15}. Only four of our galaxies have $r_\mathrm{e} \leq 0.96'' = 0.8 \times 1.2''$. They are included in the following figures, but marked as probably unresolved.

\begin{figure}
	\resizebox{\hsize}{!}{\includegraphics{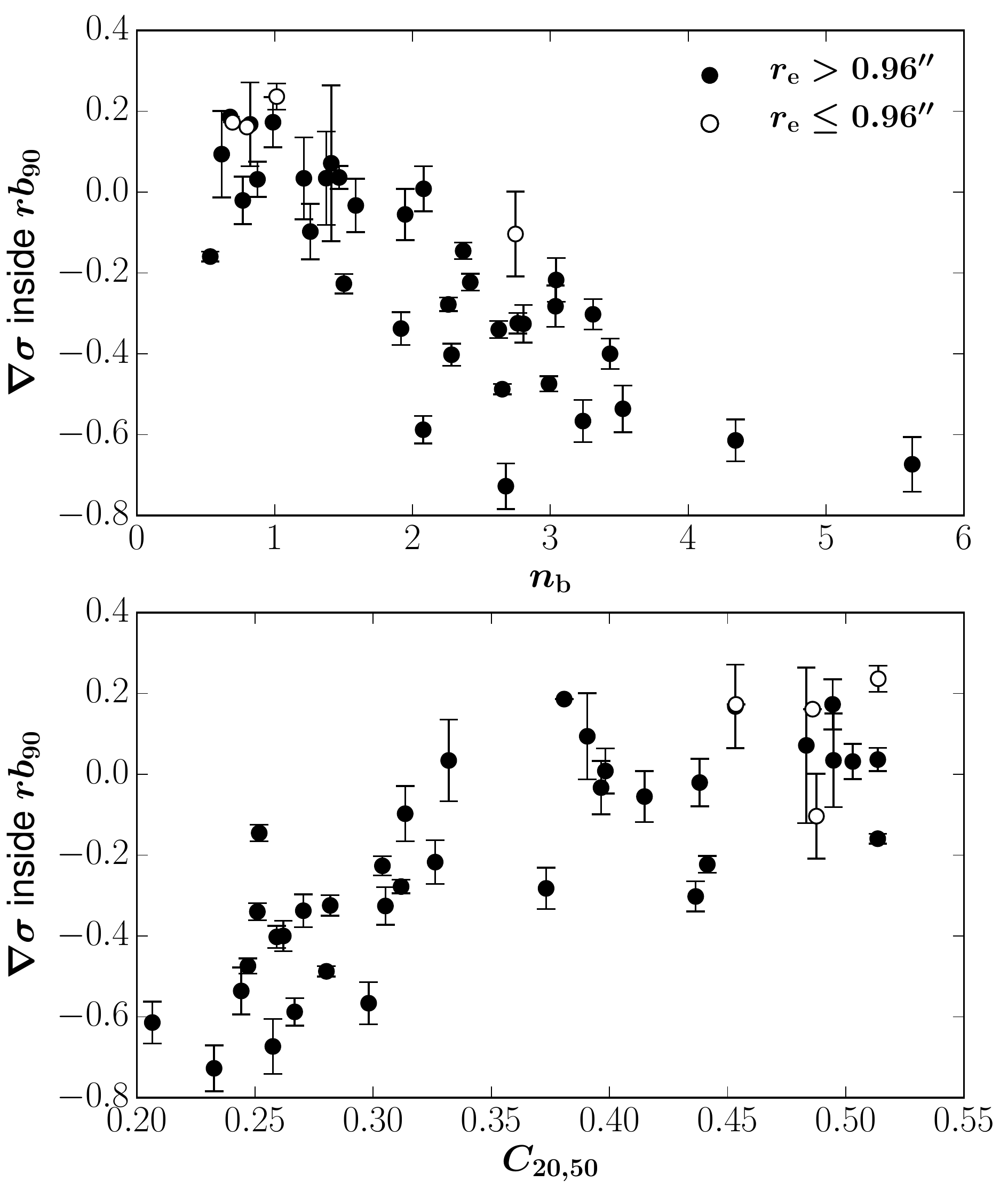}}
	\caption{The upper panel shows an anti-correlation between the velocity dispersion gradient inside the bulge radius $rb_{90}$ and the bulge Sérsic index $n_\mathrm{b}$. The lower plot shows a correlation between the velocity dispersion gradient with the concentration index $C_{20,50}$. The Spearman's rank correlation coefficient is $\rho = -0.80$  in the upper panel and $\rho = 0.81$ in the lower panel. Empty circles mark the bulges that are not well resolved. The error bars indicate the uncertainty of the fit of the velocity dispersion profile.}
	\label{sigma-slope-bulge}
\end{figure}

In Fig. \ref{sigma-slope-bulge} we present the relation between stellar velocity dispersion gradient $\nabla\sigma$ in the bulge region and bulge Sérsic index. The gradient is derived from the radial velocity dispersion profile by first normalising it to the bulge radius $rb_{90}$ and the central velocity dispersion $\sigma_0$ and then fitting a linear function to the velocity dispersion within the bulge radius. During the regression process we weight the data values by the associated uncertainties. We denote the slope of that function as $\nabla\sigma$.

We find an anti-correlation between $\nabla\sigma$ and $n_\mathrm{b}$ and a correlation between $\nabla\sigma$ and $C_{20,50}$. Low-$n_\mathrm{b}$ (high-$C_{20,50}$) bulges have approximately flat profiles whereas high-$n_\mathrm{b}$ (low-$C_{20,50}$) bulges have slopes as steep as $\nabla\sigma \approx -0.7$. This result is in good agreement with findings from \citet{Fabricius12}. They observed rather flat profiles for pseudobulges and centrally peaked profiles for classical bulges.

However, these relations must be considered carefully, since they may be significantly influenced by bulge size alone. Let's assume, e.g., that all galaxies had identical centrally peaked velocity dispersion profiles, but different bulge sizes. Small bulges would then have comparatively small values for $\nabla\sigma$ whereas for larger bulges we would measure a larger decrease in $\sigma$, solely because of the normalisation of the radial profile to the bulge radius. Hence, the distribution of bulge radii alone can theoretically produce the observed relations. As a matter of fact, pseudobulges are usually smaller than classical bulges and there is a correlation between Sérsic index and bulge size. The imprint of the bulge size on the anti-correlation between $\nabla\sigma$ and $n_\mathrm{b}$ is thus inevitable.

\begin{figure}
	\resizebox{\hsize}{!}{\includegraphics{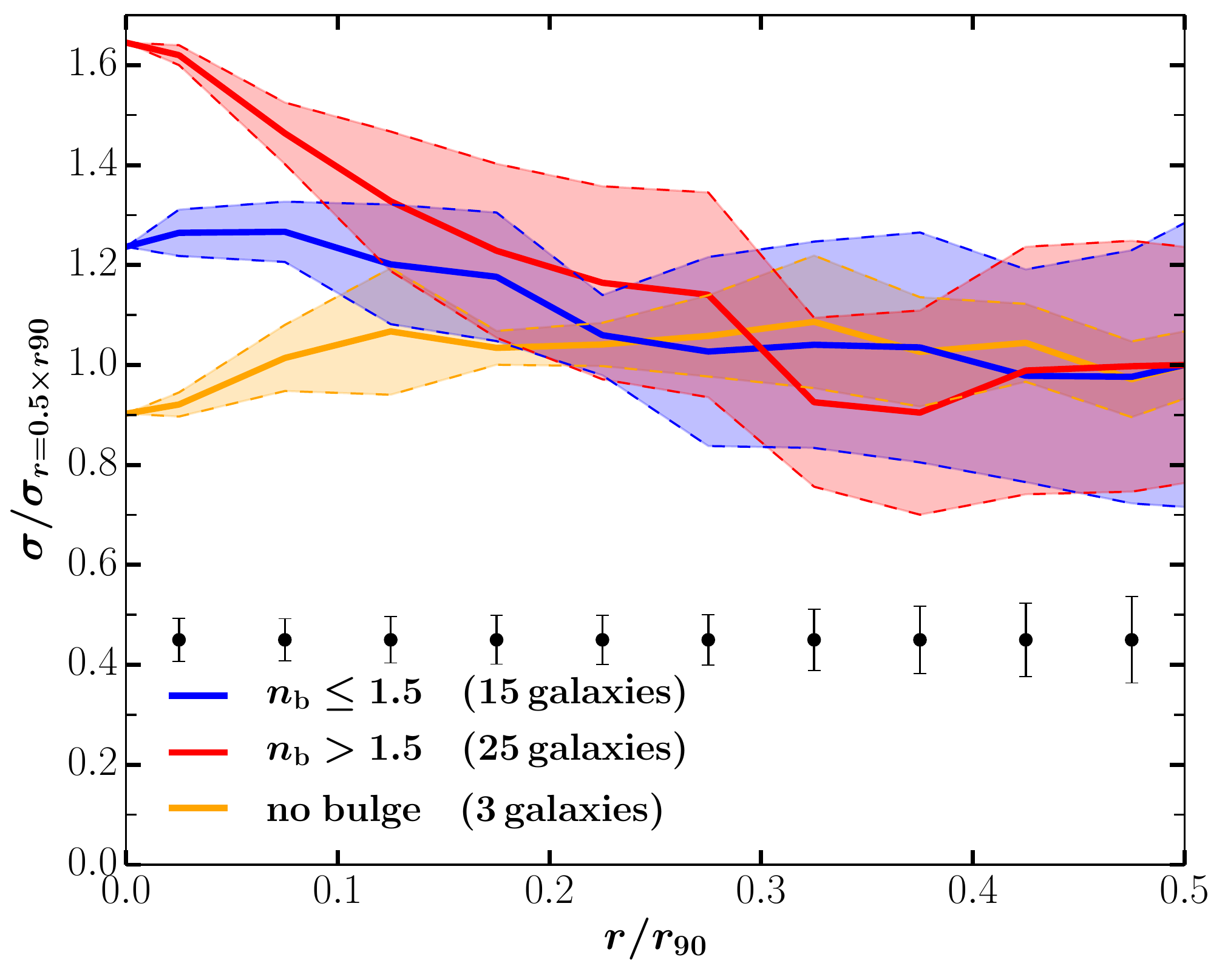}}
	\caption{Average radial velocity dispersion profiles for low-$n_\mathrm{b}$ (blue), high-$n_\mathrm{b}$ (red) and bulgeless (orange) galaxies. The thick solid lines represent the median profiles and the dashed lines the median absolut deviations. The velocity dispersion is normalised by $\sigma$ at $r=0.5 \times r_{90}$ and the radial distance is normalised by $r_{90}$ -- the radius that encloses $90\%$ of the total light of the galaxy. Error bars at the bottom indicate the median uncertainty for each $0.05\times r_{90}$ bin.}
	\label{fig:sigma-profile-r90}
\end{figure}

An alternative is to define a radius that is independent of the bulge, in which we fit the velocity dispersion profile. Fig. \ref{fig:sigma-profile-r90} shows the global radial velocity dispersion profiles for all galaxies of the sample averaged within three different groups: low-$n_\mathrm{b}$, high-$n_\mathrm{b}$ and bulgeless galaxies. The y-axis is normalised by the velocity dispersion at $0.5 \times r_{90}$ and the x-axis by the $r_{90}$ parameter derived from the growth curves, i.e. the radius that encloses $90\%$ of the \emph{total} light of the galaxy. This radius covers the major part of the galaxy and is located far outside the bulge. Note that not all galaxies have kinematic coverage up to $r_{90}$. The bulgeless galaxies show flat profiles throughout most of the radial extent. Galaxies with low-$n_\mathrm{b}$ bulges show on average profiles that increase towards the centre by $\approx 20\%$, but are close to flat in the inner $\approx 0.15 \times r_{90}$. Galaxies with high-$n_\mathrm{b}$ bulges show a stronger increase up to $\approx 60 \%$ with the steepest part in the most central region.

This figure confirms the same trend that we have seen before. The central parts of galaxies with classical bulges tend to have centrally peaked velocity dispersion profiles whereas galaxies that host pseudobulges have profiles that are rather centrally flat, partly similar to the profiles shown by bulgeless galaxies.

\begin{figure}
	\resizebox{\hsize}{!}{\includegraphics{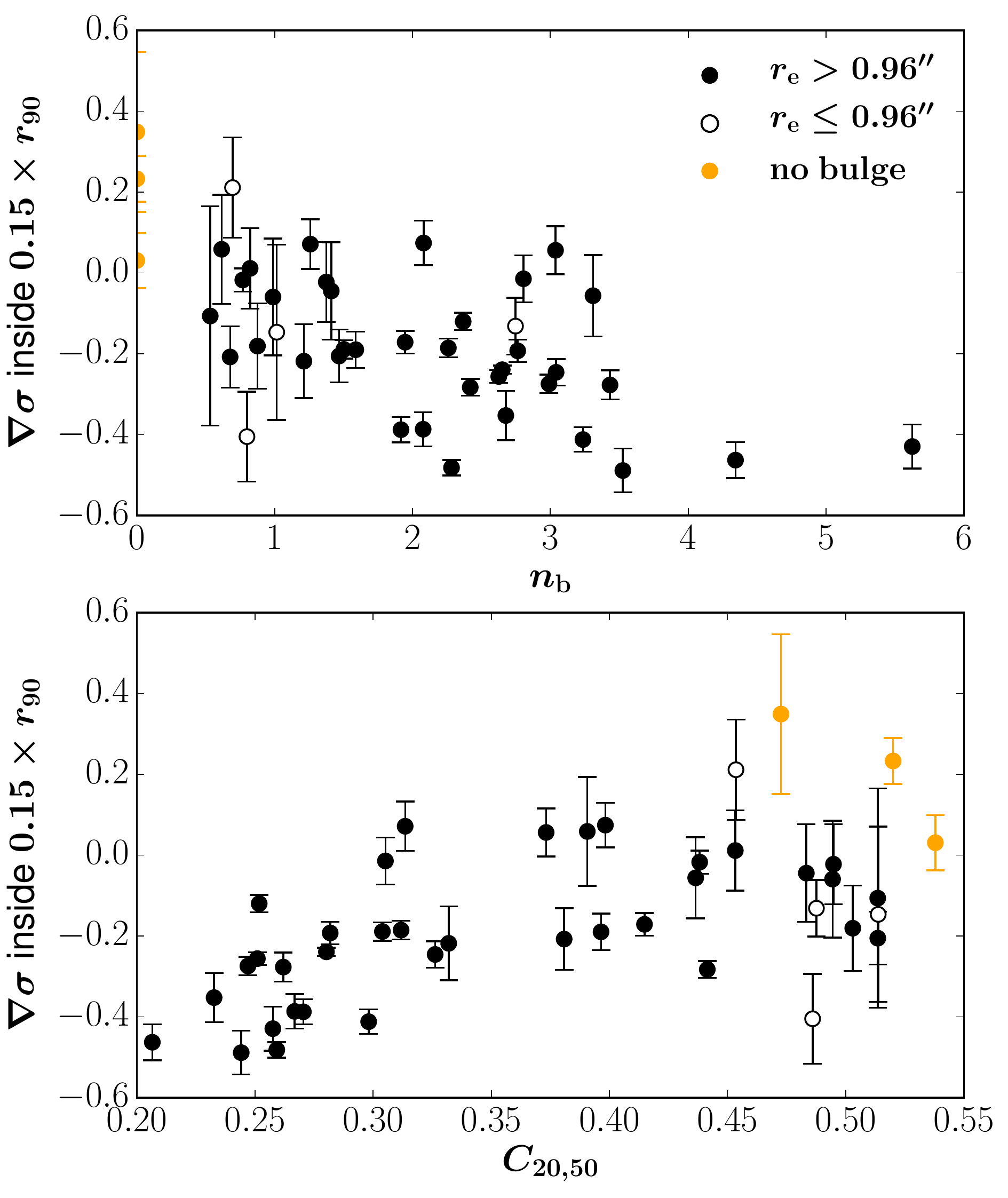}}
	\caption{The upper panel shows an anti-correlation between the velocity dispersion gradient inside $0.15 \times r_{90}$ -- the radius that encloses $90\%$ of the total light of the galaxy -- and the bulge Sérsic index $n_\mathrm{b}$. The lower plot shows a correlation between the velocity dispersion gradient with the concentration index $C_{20,50}$. The Spearman's rank correlation coefficient is $\rho = -0.53$  in the upper panel and $\rho = 0.64$ in the lower panel. Bulgeless galaxies are marked in orange. Empty circles mark the bulges that are not well resolved. The error bars indicate the uncertainty of the fit of the velocity dispersion profile.}
	\label{fig:cidx-slope-r90}
\end{figure}

In order to quantify the observed trend in the central region of the velocity dispersion profiles, the slope within $0.15 \times r_{90}$ was calculated. The choice of that radius is a compromise between not being too small and lose too much information on the larger bulges and not being too large and then contaminated by too much disc light where bulges are small. We remind the reader that we want to measure the behaviour of the velocity dispersion profile in the central region of disc galaxies, but without being affected by the bulge size. The result is shown in Fig. \ref{fig:cidx-slope-r90} as compared to $n_\mathrm{b}$ and $C_{20,50}$, respectively. We see the same trend as before, but with only a mild correlation coefficient of $\rho = -0.53$ in the upper panel and $\rho = 0.64$ in the lower panel and a fair amount of scatter that we will discuss in Sect. \ref{sect:discuss}. Furthermore, we observe that bulgeless galaxies (pure discs) behave in this figure like pseudobulges (inner discs) as expected. 

\begin{figure}
	\resizebox{\hsize}{!}{\includegraphics{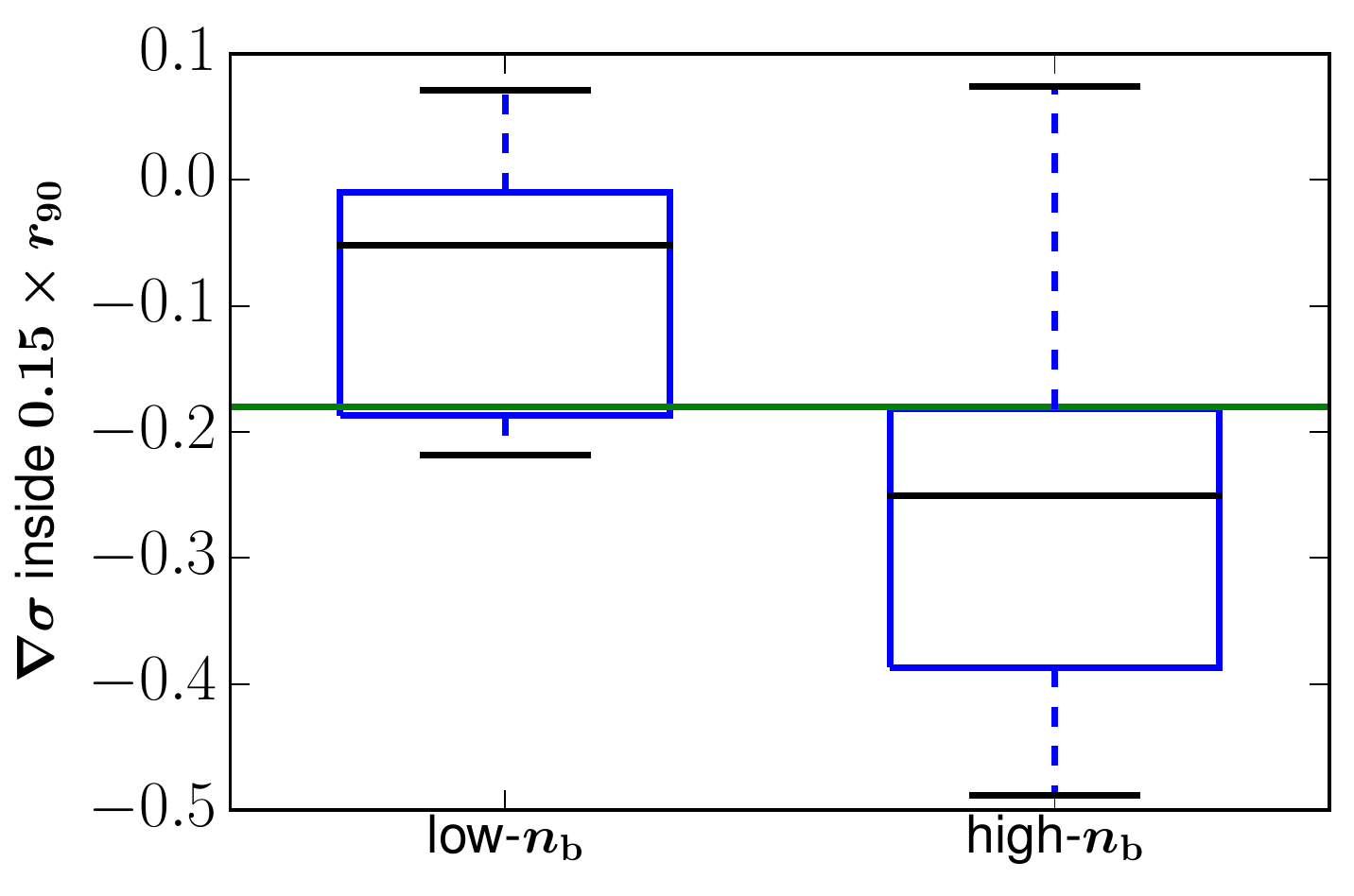}}
	\caption{Boxplot for the velocity dispersion gradient inside $0.15 \times r_{90}$ for high- and low-$n_\mathrm{b}$ galaxies. The blue box marks the interquartile range of the sample, the black line in the box gives the median value. The top and bottom black line stand for the highest and lowest value, respectively. The thick green line at $\nabla\sigma = -0.18$ gives a good demarcation to separate the subsamples based on the velocity dispersion gradient.}
	\label{fig:boxplot-slope-r90}
\end{figure}

We evaluate the results statistically further in Fig. \ref{fig:boxplot-slope-r90}, where we show a box plot for low- and high-$n_\mathrm{b}$ galaxies. The median $\nabla\sigma$ for low-$n_\mathrm{b}$ galaxies is at $-0.05$ and for high-$n_\mathrm{b}$ galaxies at $-0.25$. The interquartile ranges for both populations are separated with one lower quartile limit coinciding approximately at $-0.18$ with the other upper quartile limit. If we choose $\nabla\sigma = -0.18$ to divide the bulges into two subsamples we are essentially separating low- and high-$n_\mathrm{b}$ bulges by a completely independent method. Following this, we have established a kinematic approach to isolate pseudobulges from classical bulges that we use in combination with traditional and new photometric parameters in Sect. \ref{sect:recipe} to classify bulges.

\section{Discussion}
\label{sect:discuss}

\subsection{A recipe for separating inner discs from classical bulges}
\label{sect:recipe}

The classification of galaxy bulges into classical bulges (presumably built from violent processes such as mergers), and pseudobulges (thought to be built from dynamical instabilities in the major disc) has become a common task in extragalactic astrophysics, yet there is still no unambiguous way of doing it. While the bulge Sérsic index is probably the most frequently used criterion for bulge type diagnostics in literature, it has been shown to be prone to misclassifications in some cases. Other criteria have been proposed, but the general consensus is that no single criterion should be used alone. For an overview we refer the reader to \citet{KormendyKennicutt04}, \citet{Athanassoula05} or the very recent review by \citet{FisherDrory15}.

As mentioned in the introduction, we use the term ``pseudobulge'' to describe the discy bulges in our sample, not only for historical reasons but also because the term has been largely adopted by the astrophysical community. With this word we are referring to inner discs built from disc material through secular evolution. In this subsection we are trying to give a recipe to separate them from classical bulges. We strongly encourage the use of a combination of photometric and kinematic bulge parameters for a safe classification. In this work we analysed and compared different approaches and determined four parameters that can be used for bulge diagnostics: The bulge Sérsic index $n_\mathrm{b}$, the concentration index $C_{20,50}$, the central velocity dispersion gradient $\nabla\sigma$ and the Kormendy relation.

We decided not to use the $B/T$ light ratio as classification criterion. Considering that pseudobulges are thought to be built from disc material, they are expected to be small fractions of their host galaxies, in contrast to classical bulges which are probably relics from merger events and independent from the disc. Observations confirm that galaxies with pseudobulges have on average smaller $B/T$ light ratios, but they also show that there is a significant overlap \citep[e.g. ][]{Drory07, Fisher08, Gadotti09}. There is no physical reason for a lower limit of $B/T$ for classical bulges. Hence, the bulge-to-total light ratio can be used for reference, but it should not be included to separate inner discs from classical bulges.

Table \ref{tbl:bulges} presents our classification of 45 CALIFA galaxies. It contains all galaxies of the sample classified by our 4 different criteria. As mentioned earlier, there are 3 bulgeless galaxies within the sample and 2 galaxies have no kinematic data.

The following rules were applied to separate between pseudobulges (\emph{ps}) and classical bulges (\emph{cl}):

\begin{equation*}
\begin{aligned}
\text{Sérsic index: }&
\begin{cases}
\text{ps} & \text{if } \quad n_\mathrm{b} \leq 1.5 \\
\text{cl} & \text{if } \quad n_\mathrm{b} > 1.5 \\
\end{cases}\\
\text{Concentration index: }&
\begin{cases}
\text{ps} & \text{if } \quad C_{20,50} \geq 0.4 \\
\text{cl} & \text{if } \quad C_{20,50} < 0.4 \\
\end{cases}\\
\text{Velocity dispersion: }&
\begin{cases}
\text{ps} & \text{if } \quad \nabla\sigma \geq -0.18 \\
\text{cl} & \text{if } \quad \nabla\sigma < -0.18 \\
\end{cases}\\
\text{Kormendy relation: }&
\begin{cases}
\text{ps} & \text{if } \text{the bulge lies below and outside}\\ &\pm 2 \sigma \text{ of the relation for elliptical}\\ &\text{galaxies} \\
\text{cl} & \text{if } \text{the bulge lies within}\\ &\text{the} \pm 2 \sigma \text{ range} \\
\end{cases}
\end{aligned}
\end{equation*}

The final classification is built upon a consensus of these parameters. If 3 out of 4 criteria agree, we consider the bulge to be safely classified. The division into classical bulges and pseudobulges is an interpretation of these results based on the agreement between the parameters and their physical meaning.

Out of 42 galaxies that host bulges we could reliably classify 40 ($95\%$). Our sample contains at least 16 pseudobulges and 24 classical bulges. If we assume the 40 ``safe'' classifications to be ``correct''\footnote{Since the ``true'' physical nature of the bulges is unknown, the parameter performances that we evaluate in this section should be interpreted as relative to each other and not absolute.}, than we can state that the Kormendy relation is the best criterion by achieving 39 out of 40 correct classifications, closely followed by the concentration index $C_\mathrm{20,50}$ with 38. The velocity dispersion gradient shows the largest amount of scatter among the classifiers misclassifying 7 bulges. 

As seen in Sect. \ref{sect:result_sigma} and also discussed in Sect. \ref{sect:discuss_kine} it still remains a difficult task to measure the velocity dispersion distribution accurately. However, despite the fact that it is indeed the weakest of the four criteria in our analysis, it is still strong and yields $82 \%$ correct classifications. This means, in other words, that it was possible to successfully classify $82 \%$ of the bulges in our sample using only measurements of the velocity dispersion.

The complexity of galaxy structure and dynamics makes simple classification methods virtually impossible, but by using a combination of photometric and spectroscopic parameters we were able to successfully separate most of the classical bulges from inner discs. We provide in Table \ref{tbl:bulges} a classification for a subsample of CALIFA galaxies that can be used for future investigation. We propose the usage of our combined approach as recipe for the diagnostic and separation of galaxy bulges.

\begin{table*}[]
\caption{Overview of bulge classification.}
\label{tbl:bulges}
\centering

\begin{tabular}{c c c c c c c c}
\hline\hline
ID & NED name & $B/T$ & $n_\mathrm{b}$ & $C_\mathrm{20,50}$ & $\nabla \sigma$ & Kormendy rel & Classification\\
	&	&	& $\leq1.5$	& $\ge0.4$ & $\ge-0.18$ &  low-outlier & \\
(1) & (2) & (3) & (4) & (5) & (6) & (7) & (8) \\
\hline
2           & UGC00005   & 0.01       & ps            & ps         & cl                    & ps                 & pseudo       \\
3           & NGC7819    & 0.12       & ps            & cl          & cl                     & ps                 &       \\
6           & NGC7824    & 0.38        & cl             & cl          & cl                     & cl                 & classical        \\
8           & NGC0001    & 0.41       & cl             & cl          & cl                    & cl                  & classical      \\
20          & NGC0160    & 0.29       & cl             & cl          & cl                     & cl                  & classical       \\
31          & NGC0234    & 0.04       & ps            & ps         & ps                    & ps                 & pseudo       \\
33          & NGC0257    & 0.10       & cl             & cl          & ps                    & cl                  & classical           \\
43          & IC1683     & 0.12       & ps            & cl          & cl                     & cl                  & classical         \\
45          & NGC0496    & bulgeless   & bulgeless       & ps         & ps                    & bulgeless            & bulgeless      \\
47          & NGC0517    & 0.55       & cl             & cl          & ps                    & cl                  & classical      \\
119         & NGC1167    & 0.23       & cl             & cl          & cl                    & cl                  & classical      \\
147         & NGC2253    & 0.08       & cl             & cl          & cl                    & cl                  & classical      \\
275         & NGC2906    & 0.14       & cl             & ps         & ps                    & ps                 & pseudo      \\
277         & NGC2916    & 0.07       & cl             & ps         & cl                     & cl                  & classical       \\
311         & NGC3106    & 0.30       & cl             & cl          & cl                     & cl                  & classical      \\
489         & NGC4047    & 0.03       & cl             & ps         & ps                    & ps                & pseudo      \\
548         & NGC4470    & bulgeless   & bulgeless       & ps         & ps                    & bulgeless           & bulgeless      \\
580         & NGC4711    & 0.02       & ps            & ps         & ps                    & ps                 & pseudo      \\
607         & UGC08234   & 0.58       & cl             & cl          & cl                     & cl                  & classical      \\
611         & NGC5016    & 0.01       & ps            & ps         & ps                    & ps                 & pseudo      \\
715         & NGC5520    & 0.26       & ps            & cl          & ps                    & cl                  &             \\
748         & NGC5633    & bulgeless   & bulgeless       & ps         & ps                    & bulgeless            & bulgeless      \\
768         & NGC5732    & 0.04       & ps            & ps         & no data                & ps                 & pseudo      \\
769         & UGC09476   & 0.03       & ps            & ps         & ps                    & ps                 & pseudo      \\
778         & NGC5784    & 0.39       & cl             & cl          & ps                    & cl                  & classical      \\
821         & NGC6060    & 0.06       & ps            & ps         & ps                    & cl                  & pseudo      \\
823         & NGC6063    & 0.01       & ps            & ps         & ps                    & ps                 & pseudo      \\
826         & NGC6081    & 0.35       & cl             & cl          & cl                    & cl                  & classical      \\
836         & NGC6155    & 0.01       & ps            & ps         & ps                    & ps                 & pseudo      \\
849         & NGC6301    & 0.01       & ps            & ps         & cl                     & ps                 & pseudo      \\
850         & NGC6314    & 0.33       & cl             & cl          & cl                     & cl                  & classical      \\
856         & IC1256     & 0.03       & cl             & ps         & ps                    & ps                 & pseudo      \\
858         & UGC10905   & 0.52       & cl             & cl          & cl                     & cl                  & classical      \\
874         & NGC7025    & 0.44        & cl             & cl          & cl                     & cl                 & classical      \\
877         & UGC11717   & 0.20       & cl             & cl          & no data                & cl                  & classical      \\
886         & NGC7311    & 0.32       & cl             & cl          & cl                     & cl                  & classical      \\
889         & NGC7364    & 0.45       & cl             & cl          & ps                    & cl                  &  classical         \\
891         & UGC12224   & 0.02       & ps            & ps         & ps                    & ps                 & pseudo      \\
898         & NGC7489    & 0.02       & ps            & ps         & cl                     & ps                 & pseudo      \\
912         & NGC7623    & 0.49        & cl             & cl          & cl                     & cl                 & classical     \\
913         & NGC7625    & 0.14       & ps            & cl          & ps                    & ps                 & pseudo      \\
915         & NGC7653    & 0.27        & cl             & cl          & cl                     & cl                 & classical      \\
916         & NGC7671    & 0.33       & cl             & cl          & cl                     & cl                  & classical      \\
917         & NGC7683    & 0.57       & cl             & cl          & cl                     & cl                  & classical      \\
923         & NGC7711    & 0.47       & cl             & cl          & cl                     & cl                  & classical      \\
\hline
\end{tabular}
\tablefoot{(1) CALIFA ID, (2) NED name, (3) bulge-to-total light ratio, for reference. We list four different bulge classification criteria in column 4-7: (4) bulge Sérsic index, (5) concentration index, (6) central velocity dispersion gradient, (7) Kormendy relation. (8) Final classification: In the last column we assign each bulge a final classification if and only if at least three out of four criteria are in agreement. With ``pseudo'' we are referring to the inner disc of galaxies built through secular evolution, as explained in more detail in the text. The second line of the head of the table shows the boundaries that we have defined to demarcate pseudobulges. If a value of a specific cell in column 4-7 is within these limits it is annotated as ``ps'', otherwise it is ``cl'', which refers to ``pseudobulge'' and ``classical bulge'', respectively. The 3 bulgeless galaxies are annotated in column 3, 4 and 7. Two galaxies have no kinematic data, and thus no value in column 6.}
\end{table*}

\subsection{The acquisition of kinematic bulge parameters}
\label{sect:discuss_kine}

The traditional picture of a bulge being only a dynamically hot central component that adds an elliptical-like de Vaucouleurs light distribution to the surface brightness profile of the surrounding disc has long been shown to be obsolete. Too much discrepancy has been found between this scenario and observational evidence. A dichotomy of bulges was observed instead. For many galaxies the extra light in the central region was found to follow rather an exponential law, the geometrical appearance was flattened by rotation and disc-like structures like nuclear spirals and nuclear bars were observed. Since then, extensive work has been done and numerous bulge classification criteria have been proposed. Yet the kinematic distinctness of the bulge types could still not been quantified satisfactorily -- despite being known over more than three decades \citep{Kormendy82}. \citet{Fabricius12} showed with convincing observational evidence using long-slit spectroscopy that dynamics are indeed part of the bulge dichotomy. Similar results were obtained for a few galaxies by \citet{MendezAbreu14} with IFS data.

In this paper we tried to address this kinematic problem with a larger set of galaxies with IFS observations from the CALIFA survey. We have found similar results to those from previously mentioned works. The stellar radial velocity dispersion gradient is close to flat for pseudobulges and centrally peaked for classical bulges. We see that it is, thus, not only the light profile of a pseudobulge that shows disc similarity, but also the stellar velocity dispersion that resembles disc behaviour. The question is how to quantify what we deduce from the virtual inspection of the profiles. 

In Fig. \ref{sigma-slope-bulge} the gradient of the velocity dispersion profile calculated in the bulge region is plotted versus the bulge Sérsic index and the light concentration. A similar approach has been used by \citet{Fabricius12}. A clear correlation can be seen in that plot, but it has two major caveats: 1) Since the velocity dispersion does not follow a linear profile, the derived slope \emph{inside} the bulge region depends strongly on the bulge radius. That, on its part, is usually bigger for classical bulges and smaller for pseudobulges, albeit not exclusively. Consequently, the velocity dispersion gradient when derived exactly over the bulge extent is automatically connected to the bulge type. A parameter that is independent from the size of the bulge is desirable. 2) The bulge radii of some pseudobulges are close to the spatial resolution limit of the CALIFA data, whereas all classical bulges are big enough to be very well resolved. Hence, some parts of the bulge-dominated region of these pseudobulges are smeared out to flat profiles.

We tried to address the first problem in Fig. \ref{fig:cidx-slope-r90} by choosing another radial aperture for measuring the velocity dispersion gradient. The radial limit $0.15 \times r_{90}$ ranges between $5''$ and $13''$ -- well beyond the CALIFA PSF FWHM. Using this limit instead of either the bulge radius or e.g. $r_{20}$ has the advantage of being likely not correlated to the bulge size, since it is a fraction of the radius that captures $90 \%$ of the \emph{total} light of the galaxy. At the same time we are ensuring with this approach that we have enough resolution elements within that region.

It is important to keep in mind that Fig. \ref{fig:cidx-slope-r90} tells us about the behaviour of the velocity dispersion profile in the central part of the galaxy and not specifically in the bulge region. The different coverage of the bulge-disc regions -- that we caused intentionally -- could introduce some additional scatter in that relation.

Pseudobulges are expected to be small objects with low velocity dispersion. Hence, in order to identify pseudobulges reliably, high spectral and spatial resolution is required. Despite the great advantage of IFS observations from surveys like MaNGA \citep[Mapping Nearby Galaxies at
APO,][]{Bundy14} or SAMI \citep[Sydney-Australian-Astronomical-Observatory
Multi-object Integral-Field Spectrograph survey,][]{Croom12} to have spectral information over a two-dimensional area on the sky, there is still a deficiency of spatial resolution as compared to data from photometric surveys. CALIFA has the advantage over SAMI and MaNGA to provide a better physical spatial resolution given the lower redshifts by similar projected resolution. As shown in this work, it allowed for a classification of $82 \%$ of the bulges using exclusively kinematics.

A separation of bulge types based on their kinematics would physically be a quite convincing approach, since pseudobulges presumably being essentially inner discs should resemble the behaviour of the surrounding discs whereas classical bulges should be observed as hot elliptical-like components. It is therefore desirable to use IFU instruments with even larger spatial resolution such as MUSE \citep[Multi-Unit Spectroscopic Explorer,][]{Bacon10} to further decrease the uncertainties of kinematic measurements of bulges.

\subsection{Introducing a new concentration index: $C_{20,50}$}

The radial light distribution of a galaxy disc is best described by either a single-exponential (Type I) or double-exponential (Type II and Type III) profile. Any additional baryonic component adds light to this distribution. The presence of a bulge can thus be observed by an excess of light in the central part of the galaxy. Classical bulges are usually -- but not exclusively -- more luminous and in itself more concentrated than pseudobulges. It might seem obvious that one of the first things to do in order to identify and classify bulges is to measure the concentration. The result, however, depends strongly on the method and parametrisation. It is possible to define a concentration index as 1) ratio between two radii that enclose certain percentages of the total light of the galaxy \citep[e.g. $r_{50}/r_{25}$, $r_{75}/r_{50}$, $r_{75}/r_{25}$, $r_{80}/r_{20}$,][]{deVaucouleurs77a, Kent85} or as 2) ratio of the flux between two correlated isophotes \citep{Doi93, Abraham94, Trujillo01}. The final value will also depend on whether the flux was measured within a Petrosian aperture or within the complete extent on the galaxy based on growth curve analysis, as shown in Appendix \ref{apx:petro}.

In this work we have demonstrated the capability of the concentration index defined as $C_{20,50} = r_{20}/r_{50}$ to diagnose the bulge type with great accuracy: 38 out of 40 correct classifications following the recipe in Sect. \ref{sect:recipe}. We have shown a strong correlation with the logarithm of the global Sérsic index of the galaxy: Spearman's rank correlation coefficient $\rho = -0.80$, and we found correlations with bulge Sérsic index, bulge-to-disc light ratio, Kormendy relation and velocity dispersion gradient. Based on these results, it is evident that $C_{20,50}$ is a powerful indicator of the bulge nature. We thus encourage the use of this index over the widely used $r_{90}/r_{50}$ as it seems to track a more bulge related part of the light distribution of the galaxy.

\section{Conclusions}
\label{Sect:Conclusions}

In this paper, we derived a set of different photometric and spectroscopic parameters that can be used to separate inner discs from bulges. For that purpose we used growth curve measurements and performed detailed 2D image decomposition of SDSS $r$-band images and spectral fitting to CALIFA IFU data cubes.

We demonstrated that the radial velocity dispersion profile of galaxies can be used to discriminate between pseudobulges and classical bulges. We cautioned on using the bulge radius, which is not independent from the bulge type, to normalise the radial profile. Instead, we found a different quantification of the velocity dispersion based on the global profile that can be used to classify bulges.

We promote the concentration index, defined as $C_{20,50} = r_{20}/r_{50}$, the ratio of the radii that enclose $20\%$ and $50\%$ of the total light of the galaxy, respectively. It correlates well with the widely used bulge Sérsic index $n_\mathrm{b}$ and yields statistically similar results when used for bulge classification. We encourage the usage of $C_{20,50}$ given that it is a parameter that can be derived with very little effort.

We showed that the concentration index $C_{20,50}$ and the Kormendy relation are the best classification criteria by achieving over $95 \%$ correct classifications (based on the agreement with the other criteria) following our recipe in Sect. \ref{sect:recipe}. When used in combination, these two criteria should yield a robust indication of the nature of bulges. However, it is important to remember that none of the criteria can undoubtfully separate bulges from inner discs. The more criteria are used, the safer the classification becomes.

We propose a recipe based on four parameters from photometry and spectroscopy to classify bulges. The different parameters are in good agreement and allow a safe classification for approximately $95\%$ of the galaxies.

By using this recipe we provided a detailed bulge classification for a subsample of 45 galaxies from the CALIFA survey. Future IFU surveys should be used to further increase the accuracy and reliability of spectroscopic analyses that are of great importance to unveil the true nature of bulges.

\begin{acknowledgements}

This study makes uses of the data provided by the Calar Alto Legacy Integral Field Area (CALIFA) survey (http://www.califa.caha.es). Based on observations collected at the Centro Astronómico Hispano Alemán (CAHA) at Calar Alto, operated jointly by the Max-Planck-Institut fur Astronomie and the Instituto de Astrofísica de Andalucía (CSIC). CALIFA is the first legacy survey being performed at Calar Alto.The CALIFA collaboration would like to thank the IAA-CSIC and MPIA-MPG as major partners of the observatory, and CAHA itself, for the unique access to telescope time and support in manpower and infrastructures. The CALIFA collaboration thanks also the CAHA staff for the dedication to this project. We thank the anonymous referee for a careful reading and several comments that helped to improve the paper. RGB and RGD acknowledge support from grants AYA2014-57490-P and JA-FQM-2828. RAM acknowledges support by the Swiss National Science Foundation. IM acknowledges financial support from grants AYA2013-42227-P and AYA2016-76682-C3-1-P. SFS thanks the CONACYT-125180, DGAPA-IA100815 and DGAPA-IA101217 projects for providing him support in this study. 

\end{acknowledgements}

\bibliographystyle{aa}
\bibliography{Bulges_in_CALIFA_2c_JN.bib}

\begin{appendix}
\section{Relation between the Petrosian concentration index and our $C_{20,50}$}
\label{apx:petro}

\begin{figure}
	\resizebox{\hsize}{!}{\includegraphics{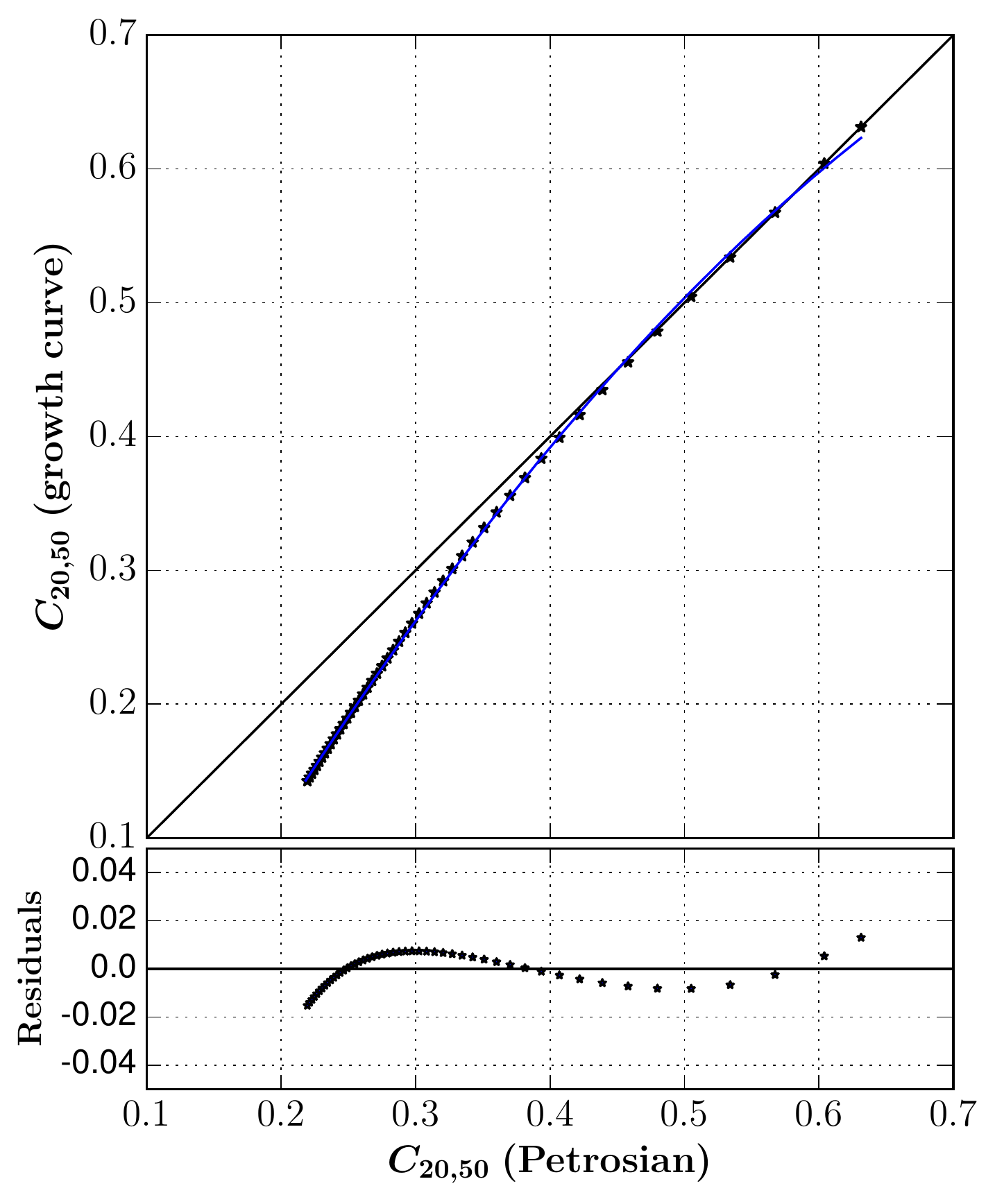}}
	\caption{Comparison between the concentration index $C_{20,50}$ from growth curve measurement with the associated Petrosian concentration index within the Petrosian aperture $2R_\mathrm{p}$. The solid blue line is a polynomial fit to the data, residuals are shown in the bottom panel. The best fit parameters are given in the text.}
	\label{fig:petro}
\end{figure}

The Petrosian radius $R_{\mathrm{p}}$ is defined to be the radius where the Petrosian function $\eta(R)$ equals some fixed value. The Petrosian function gives the average intensity within some projected radius divided by the intensity at that radius. Different multiples of $R_{\mathrm{p}}$ have been used to measure galaxy magnitudes, but they generally underestimate the total flux of the galaxy. Thus any radius $r_k$, where $r_k$ encloses $k$ percent of the total flux, is also underestimated. \citet{Graham05} offers correction factors for magnitudes, half-light radii and surface brightness. To provide such a correction factor one has to assume a specific surface brightness distribution of the galaxies. The easiest approach is a single Sérsic profile, which as we know is a precarious simplification for most galaxies. If we adopt from the SDSS consortium the practice of measuring the flux within $2 \times R_\mathrm{p}$ and $1/\eta(R_\mathrm{p}) = 0.2$, we can directly integrate the light profiles and derive the concentration index as a function of $n$. The discrepancy between the Petrosian concentration index and the concentration derived from integrating the Sérsic function to infinity is illustrated in Fig. \ref{fig:petro}. The conversion factor between our concentration index $C_{20,50}$ and the associated Petrosian concentration can be approximated by

\begin{align}
	C_{20,50}\,\text{(gc)} = a_0+a_1\,C_{20,50}\,\text{(petro)}+a_2\,C_{20,50}\,\text{(petro)}^2
\end{align}

where $a_0=-0.23$, $a_1=1.92$ and $a_2=-0.90$. We emphasise that this approximation is under the assumption that the light profile is well described by a Sérsic function and that the ``edge'' of the galaxy is accurately derived from the growth curve measurement.

\section{Parameters of the CALIFA subsample used in this work}
\label{apx:alldata}

\begin{table*}[]
\caption{Summary of sample parameters.}
\label{tbl:alldata}
\centering

\begin{tabular}{c c c c c c c c c c c c}
\hline\hline
ID & NED name & $n_\mathrm{g}$ & $C_\mathrm{20,50}$ & $B/T$ & $n_\mathrm{b}$ & $r_\mathrm{e}$ & $\langle\mu_\mathrm{e}\rangle$ & $rb_{90}$ & $0.15\times r_\mathrm{90}$ & $\sigma_0$ & $\nabla\sigma$ \\
(1) & (2) & (3) & (4) & (5) & (6) & (7) & (8) & (9) & (10) & (11) & (12)\\
\hline
2 & UGC00005 & 0.79 & 0.50 & 0.01 & 0.88 & 1.22 & 19.15 & 2.68 & 5.40 & 99.2 $\pm$ 4.1 & -0.18 $\pm$ 0.11\\
3 & NGC7819 & 7.25 & 0.33 & 0.12 & 1.21 & 1.95 & 18.93  & 4.93 & 8.90 & 88.4 $\pm$ 5.0 & -0.22 $\pm$ 0.09\\
6 & NGC7824 & 5.89 & 0.23 & 0.38 & 2.68 & 3.66 & 17.97  & 14.71 & 6.41 & 212.8 $\pm$ 2.7 & -0.35 $\pm$ 0.06\\
8 & NGC0001 & 5.27 & 0.28 & 0.41 & 2.76 & 3.48 & 18.15  & 14.30 & 7.98 & 120.7 $\pm$ 3.4 & -0.19 $\pm$ 0.03\\
20 & NGC0160 & 8.05 & 0.25 & 0.29 & 2.99 & 5.62 & 18.74  & 24.52 & 11.36 & 185.3 $\pm$ 2.2 & -0.27 $\pm$ 0.02\\
31 & NGC0234 & 0.91 & 0.45 & 0.04 & 0.82 & 1.62 & 18.62  & 3.50 & 6.78 & 74.7 $\pm$ 4.0 & 0.01 $\pm$ 0.10\\
33 & NGC0257 & 2.97 & 0.40 & 0.10 & 2.08 & 2.92 & 18.53  & 9.92 & 7.82 & 94.3 $\pm$ 2.3 & 0.07 $\pm$ 0.05\\
43 & IC1683 & 3.75 & 0.38 & 0.12 & 0.68 & 0.97 & 17.07  & 1.95 & 7.48 & 99.7 $\pm$ 8.2 & -0.21 $\pm$ 0.08\\
45 & NGC0496 & 1.02 & 0.47 &  &  &  &  &  & 6.05 & 58.3 $\pm$ 11.2 & 0.35 $\pm$ 0.20\\
47 & NGC0517 & 5.61 & 0.37 & 0.55 & 3.04 & 4.31 & 17.70 & 19.05 & 8.20 & 145.7 $\pm$ 5.4 & 0.06 $\pm$ 0.06\\
119 & NGC1167 & 4.83 & 0.31 & 0.23 & 2.26 & 5.18 & 18.90 & 18.53 & 12.07 & 188.8 $\pm$ 5.4 & -0.19 $\pm$ 0.02\\
147 & NGC2253 & 2.10 & 0.40 & 0.08 & 1.59 & 1.18 & 17.43 & 3.43 & 6.32 & 101.4 $\pm$ 2.9 & -0.19 $\pm$ 0.04\\
275 & NGC2906 & 1.36 & 0.44 & 0.14 & 3.31 & 4.26 & 18.80  & 20.15 & 6.93 & 83.2 $\pm$ 3.1 & -0.06 $\pm$ 0.10\\
277 & NGC2916 & 1.30 & 0.44 & 0.07 & 2.42 & 2.16 & 18.03  & 8.09 & 9.83 & 107.9 $\pm$ 3.8 & -0.28 $\pm$ 0.02\\
311 & NGC3106 & 12.06 & 0.24 & 0.30 & 3.53 & 3.60 & 18.37  & 17.94 & 12.77 & 171.6 $\pm$ 4.0 & -0.49 $\pm$ 0.05\\
489 & NGC4047 & 1.82 & 0.41 & 0.03 & 1.95 & 1.00 & 18.18  & 3.24 & 8.89 & 91.8 $\pm$ 2.3 & -0.17 $\pm$ 0.03\\
548 & NGC4470 & 0.59 & 0.54 &  &  &  &  &  & 5.98 & 50.3 $\pm$ 4.6 & 0.03 $\pm$ 0.07\\
580 & NGC4711 & 0.82 & 0.49 & 0.02 & 0.99 & 0.97 & 18.88  & 2.23 & 5.80 & 56.6 $\pm$ 5.2 & -0.06 $\pm$ 0.14\\
607 & UGC08234 & 6.37 & 0.26 & 0.58 & 5.68 & 5.08 & 18.41  & 39.45 & 7.76 & 197.5 $\pm$ 2.5 & -0.43 $\pm$ 0.05\\
611 & NGC5016 & 1.32 & 0.45 & 0.01 & 0.69 & 0.95 & 18.88  & 1.92 & 7.63 & 54.9 $\pm$ 3.7 & 0.21 $\pm$ 0.12\\
715 & NGC5520 & 3.02 & 0.31 & 0.26 & 1.26 & 3.59 & 18.05 7 & 9.25 & 6.26 & 85.7 $\pm$ 1.8 & 0.07 $\pm$ 0.06\\
748 & NGC5633 & 0.85 & 0.52 &  &  &  &  &  & 6.04 & 64.0 $\pm$ 3.1 & 0.23 $\pm$ 0.06\\
768 & NGC5732 & 1.66 & 0.44 & 0.04 & 0.99 & 1.49 & 19.42  & 3.45 & 5.22 &  &  \\
769 & UGC09476 & 1.00 & 0.49 & 0.03 & 1.37 & 2.12 & 19.87  & 5.69 & 6.73 & 50.5 $\pm$ 3.9 & -0.02 $\pm$ 0.10\\
778 & NGC5784 & 5.56 & 0.25 & 0.39 & 2.37 & 4.04 & 17.84  & 14.91 & 10.4 & 209.7 $\pm$ 3.1 & -0.12 $\pm$ 0.02\\
821 & NGC6060 & 2.76 & 0.44 & 0.06 & 0.77 & 2.77 & 18.42  & 5.84 & 9.37 & 112.3 $\pm$ 4.8 & -0.02 $\pm$ 0.03\\
823 & NGC6063 & 0.66 & 0.51 & 0.01 & 0.97 & 0.92 & 19.58  & 2.01 & 6.15 & 48.2 $\pm$ 6.2 & -0.15 $\pm$ 0.22\\
826 & NGC6081 & 3.79 & 0.30 & 0.35 & 1.50 & 3.22 & 18.14  & 9.07 & 6.51 & 194.9 $\pm$ 3.4 & -0.19 $\pm$ 0.02\\
836 & NGC6155 & 0.95 & 0.51 & 0.01 & 0.53 & 1.11 & 18.73  & 2.09 & 5.82 & 78.0 $\pm$ 16.2 & -0.11 $\pm$ 0.27\\
849 & NGC6301 & 0.77 & 0.51 & 0.01 & 1.47 & 1.12 & 19.20  & 3.12 & 7.64 & 80.7 $\pm$ 5.1 & -0.21 $\pm$ 0.07\\
850 & NGC6314 & 6.20 & 0.25 & 0.33 & 2.63 & 2.99 & 17.48  & 11.86 & 7.72 & 174.1 $\pm$ 1.8 & -0.26 $\pm$ 0.02\\
856 & IC1256 & 1.28 & 0.49 & 0.03 & 2.75 & 0.73 & 18.18  & 3.01 & 6.31 & 81.5 $\pm$ 6.3 & -0.13$\pm$ 0.07\\
858 & UGC10905 & 10.35 & 0.21 & 0.52 & 4.34 & 5.32 & 18.45  & 31.83 & 12.98 & 221.4 $\pm$ 2.5 & -0.46 $\pm$ 0.04\\
874 & NGC7025 & 5.32 & 0.28 & 0.44 & 2.65 & 6.60 & 18.48  & 26.36 & 11.34 & 225.2 $\pm$ 2.1 & -0.24 $\pm$ 0.01\\
877 & UGC11717 & 4.35 & 0.35 & 0.20 & 2.53 & 2.97 & 18.81  & 11.46 & 8.01 &  & \\
886 & NGC7311 & 4.29 & 0.27 & 0.32 & 2.22 & 2.75 & 17.19 & 6.48 & 7.33 & 184.1 $\pm$ 2.7 & -0.39 $\pm$ 0.03\\
889 & NGC7364 & 3.77 & 0.31 & 0.45 & 2.80 & 5.59 & 18.71  & 23.31 & 9.54 & 132.4 $\pm$ 3.3 & -0.01 $\pm$ 0.06\\
891 & UGC12224 & 0.98 & 0.48 & 0.02 & 1.49 & 2.11 & 20.25  & 5.37 & 8.18 & 59.8 $\pm$ 9.6 & -0.04 $\pm$ 0.12\\
898 & NGC7489 & 1.26 & 0.49 & 0.02 & 0.80 & 0.79 & 19.09  & 1.69 & 7.22 & 85.6 $\pm$ 7.0 & -0.40 $\pm$ 0.11\\
912 & NGC7623 & 4.42 & 0.27 & 0.49 & 2.08 & 4.50 & 17.99  & 15.25 & 7.86 & 165.7 $\pm$ 1.5 & -0.39 $\pm$ 0.04\\
913 & NGC7625 & 1.77 & 0.39 & 0.14 & 0.62 & 4.26 & 18.67  & 8.33 & 6.03 & 70.6 $\pm$ 2.9 & 0.06 $\pm$ 0.13\\
915 & NGC7653 & 2.40 & 0.33 & 0.27 & 3.04 & 3.72 & 18.78  & 16.43 & 5.80 & 100.6 $\pm$ 1.6 & -0.25 $\pm$ 0.03\\
916 & NGC7671 & 6.10 & 0.26 & 0.33 & 2.12 & 1.77 & 16.71  & 6.69 & 8.26 & 248.2 $\pm$ 2.1 & -0.48 $\pm$ 0.02\\
917 & NGC7683 & 4.18 & 0.30 & 0.57 & 3.24 & 7.59 & 18.76  & 35.24 & 10.15 & 214.0 $\pm$ 5.5 & -0.41 $\pm$ 0.03\\
923 & NGC7711 & 5.89 & 0.26 & 0.47 & 3.38 & 4.94 & 18.09  & 24.50 & 10.80 & 186.1 $\pm$ 1.0 & -0.28 $\pm$ 0.04\\
\hline
\end{tabular}
\tablefoot{(1) CALIFA ID, (2) NED name, (3) global Sérsic index, (4) concentration index, (5) bulge-to-total light ratio, (6) bulge Sérsic index, (7) effective radius in arcsec,  (8) mean effective surface brightness in mag $\mathrm{arcsec^{-2}}$, (9) bulge radius as defined in Sect. \ref{sect:bulgeradius} in arcsec, (10) 0.15 $\times$ the radius that encloses $90 \%$ of the total light of the galaxy in arcsec, (11) central velocity dispersion in km $\mathrm{s^{-1}}$, (12) velocity dispersion gradient.}
\end{table*}

\end{appendix}

\end{document}